\documentclass[aps,prd,preprint,nofootinbib,superscriptaddress,floatfix]{revtex4-2}
\usepackage{amsmath,amssymb,bm,mathtools,mathrsfs}
\usepackage{graphicx}
\usepackage{booktabs}
\usepackage{hyperref}
\usepackage{xcolor}
\usepackage{microtype}
\usepackage{enumitem}
\usepackage{siunitx}
\hypersetup{colorlinks=true,linkcolor=blue,citecolor=blue,urlcolor=blue}

\makeatletter
\def\frontmatter@abstract@produce{%
  \par
  \addvspace{\frontmatter@preabstractspace}%
  \begingroup
    \dimen@\baselineskip
    \setbox\z@\vtop{\unvcopy\absbox}%
    \advance\dimen@-\ht\z@\advance\dimen@-\prevdepth
    \@ifdim{\dimen@>\z@}{\vskip\dimen@}{}%
  \endgroup
  \begingroup
    \prep@absbox
    \unvbox\absbox
    \post@absbox
  \endgroup
  \@ifx{\@empty\mini@notes}{}{\mini@notes\par}%
  \addvspace\frontmatter@postabstractspace
}%
\makeatother
\graphicspath{{figures/}}

\newcommand{\dd}{\mathrm d}
\newcommand{\rhob}{\rho_B}
\newcommand{\rhodm}{\rho_{\rm DM}}

\newcommand{\Fspin}{\mathcal F}
\newcommand{\Psic}{\Psi}
\newcommand{\Bcrit}{\mathcal B}
\newcommand{\Mcal}{\mathcal M}

\newcommand{\Ncal}{\mathcal N}
\newcommand{\Dcal}{\mathcal D}
\newcommand{\Gcal}{\mathcal G}

\begin{document}

\title{A Self-Interacting Dark-Matter Route to Supermassive Kerr Black Holes via Super-Eddington Bondi Accretion: Mass-Spin Evolution}

\author{Chian-Shu Chen}
\email{chianshu@gmail.com}
\affiliation{Department of Physics, Tamkang University, New Taipei 251, Taiwan}
\affiliation{Physics Division, National Center for Theoretical Sciences, Taipei 10617, Taiwan}
\author{Feng-Li Lin}
\email{fengli.lin@gmail.com}
\affiliation{Department of Physics, National Taiwan Normal University, Taipei 11677, Taiwan}

\date{October 2026}

\begin{abstract}
Recent JWST observations motivate mechanisms for rapid black-hole growth in the early Universe. Extending our previous Schwarzschild study, we show that critical relativistic Bondi accretion of repulsively self-interacting dark matter remains an efficient super-Eddington channel for Kerr black holes with rotating inflow. Within a stationary equatorial quasi-spherical framework, smooth critical solutions exist over a finite range of inflow angular momentum set by the Kerr marginally bound capture values. Rotation changes the sonic structure and accretion efficiency, while the critical density cancellation and dark-sector/halo matching retain their Schwarzschild form. The resulting rapid growth persists until the finite halo supply becomes limiting. Because the inflow also carries angular momentum, stellar-mass seeds can grow and spin up simultaneously; illustrative population evolutions therefore provide a joint mass--spin route to early supermassive Kerr black holes.
\end{abstract}

\maketitle
\clearpage
\tableofcontents

\section{Introduction}\label{sec:intro}

The James Webb Space Telescope (JWST) has changed the empirical landscape of early black-hole growth.  Deep spectroscopy has uncovered faint broad-line active galactic nuclei (AGNs) at $z\gtrsim4$ in numbers well above pre-JWST expectations \cite{Harikane2023,Matthee2024,Lambrides2026}, while individual systems push the problem to still earlier epochs.  CEERS\_1019 hosts an accreting black hole at $z=8.679$ only about $570\,$Myr after the Big Bang \cite{Larson2023}; GN-z11 at $z=10.6$ shows evidence for an actively accreting black hole \cite{Maiolino2024}; and the JWST-identified galaxy UHZ1, combined with its Chandra X-ray counterpart, points to a massive black hole at $z\simeq10.3$ \cite{Bogdan2024}.  At later times, JADES has revealed a dormant but extremely overmassive black hole at $z=6.68$, with a black-hole mass comparable to a substantial fraction of its host's stellar mass \cite{Juodzbalis2024}.  Taken together with the emerging population of compact ``little red dots'' and other broad-line AGN candidates, these observations sharpen the long-standing question of how black holes can reach large masses within the first billion years of cosmic history.  The interpretation of individual sources is still affected by virial-mass systematics, selection effects, obscuration, and the possibility of super-Eddington phases, but the enlarged high-redshift census has made rapid black-hole assembly a central problem rather than an exceptional one \cite{Volonteri2010,Inayoshi2020,Lambrides2026}.

Several astrophysical pathways have been proposed to explain the rapid assembly of SMBHs at high redshift. These include massive seeds formed through direct gas collapse \cite{Bromm:2002hb, Begelman:2006db}, lighter seeds originating from Population III stellar remnants \cite{2002RvMP...74.1015W}, and subsequent rapid growth through sustained or episodic super-Eddington baryonic accretion \cite{Madau:2014pta}. Hierarchical black-hole mergers may provide an additional contribution to mass assembly \cite{Volonteri:2002vz}. Primordial black holes constitute another possible class of early seeds \cite{Carr:2020xqk}. These scenarios rely on different assumptions about seed formation, accretion efficiency, and available growth time \cite{Inayoshi:2019fun}. In our previous work \cite{ChenLin2026}, we proposed a complementary nonbaryonic route in which stellar-mass primordial black holes (PBHs) grow through relativistic Bondi accretion of self-interacting dark matter (SIDM).  For a repulsively self-interacting scalar with a quartic interaction, the relativistic fluid admits a critical transonic regime in which the apparent pole of the Bondi flux is canceled by the simultaneously vanishing asymptotic SIDM density. When we use the finite halo as the outer boundary condition, the resulting critical growth coefficient is controlled primarily by the microscopic SIDM scale rather than the local halo density.  The corresponding $M^2$ growth can therefore drive stellar-mass seeds toward the supermassive regime on cosmological timescales, subject ultimately to the finite SIDM supply.  This mechanism differs qualitatively from super-Eddington baryonic growth: it does not rely on radiative inefficiency to evade the Eddington limit, and its characteristic rate instead depends on the dark-sector equation of state and self-scattering parameters.

A Kerr generalization is required to make this formation channel dynamically complete.  Astrophysical black holes generically carry angular momentum, and the accreting SIDM need not carry the same angular momentum per unit energy as the hole. In Kerr spacetime, frame dragging and the inflow's angular-momentum content modify the relativistic flow structure, potentially shifting the sonic point and altering the critical accretion rate. To distinguish these effects, we characterize the black hole by its mass $M$ and angular momentum $J$, and the fluid by its conserved angular momentum per unit accreted energy $L$. The corresponding dimensionless parameters are $\chi\equiv J/M^2$ and $\ell\equiv L/M$, respectively. We then express the Kerr metric in terms of $J/M=\chi M$, without introducing an additional dimensional spin parameter. This two-parameter formulation lets us examine the existence and properties of critical SIDM accretion solutions across the $(\chi,\ell)$ parameter space and investigate how the associated energy and angular-momentum fluxes determine the coupled evolution of black-hole mass and spin.

Relativistic accretion in Kerr spacetime has, of course, a substantial literature, but most of it addresses questions different from the SIDM critical-growth problem considered here.  Bondi's spherical solution and Michel's relativistic Schwarzschild flow provide the basic transonic framework \cite{Bondi1952,Michel1972}.  For Kerr black holes, Pariev studied Bondi-type adiabatic accretion with vanishing fluid angular momentum and obtained spin-dependent corrections to the sonic surface and rate in a cold-flow expansion \cite{Pariev1996}.  Relativistic Bondi--Hoyle calculations examined non-axisymmetric wind accretion and the spin-dependent morphology and momentum fluxes of the flow \cite{Font1999}.  A separate line of work developed stationary and transonic Kerr disks, including viscous horizon-penetrating solutions and their sonic boundary conditions \cite{PeitzAppl1997,Takahashi2007}.  Low-angular-momentum inviscid flows have been analyzed as autonomous dynamical systems and in multi-transonic settings, where the Kerr spin, prograde or retrograde angular momentum, shocks, and centrifugal barriers can qualitatively change the global solution \cite{Chakrabarti1996,Goswami2007,Das2012,ShaikhDas2018,DasNandi2019,Sukova2017}.  Multidimensional simulations likewise show that nominally Bondi-like flows can reorganize into thick equatorial structures, polar channels, shocks, or inflow--outflow configurations \cite{Mach2018,Aguayo2021,Dihingia2025}.

Our perspective differs in three related ways.  First, the fluid is not an ordinary baryonic polytrope but the hydrodynamic limit of a repulsively self-interacting scalar, whose equation of state produces the special critical cancellation identified in our previous work. Second, rather than fixing either the black-hole spin or the fluid angular momentum as the sole rotational parameter, we keep the pair $(\chi,\ell)$ and ask for the complete domain of critical SIDM solutions.  Third, we follow both the energy and angular-momentum fluxes into the hole and use them to connect the stationary flow problem to the cosmological evolution of the black-hole mass and spin.  We work in an equatorial quasi-spherical, stationary, inviscid sector with an effective $4\pi$ flux normalization.  This intentionally isolates the relativistic critical mechanism; it is not intended to replace the multidimensional disk, shock, or transport physics that becomes important outside this analytic sector.

The generic analysis gives a simple answer to the existence question.  The critical condition reduces to a quartic equation for the sonic radius, and the physical outer branch exists over a finite interval in $\ell$ whose endpoints coincide with the prograde and retrograde Kerr marginally bound angular momenta. Throughout this domain, the critical SIDM density cancellation found in Schwarzschild spacetime persists, while a calculable geometric factor $\Psi(\chi,\ell)$ multiplies the universal rate.  This makes the effect of rotation intrinsically two-dimensional in parameter space: the black-hole spin alone does not determine the rate.  In particular, the exact $\ell=0$ branch has a spin-dependent sonic radius but retains the Schwarzschild critical rate.  The matched corotating family $L=J/M$ (equivalently $\ell=\chi$) is another analytic slice of the generic solution and provides useful closed-form checks, but it is not the generic organizing principle of the Kerr problem.

The angular-momentum flux introduces a second layer of phenomenology.  Since $\dot J/\dot M_{\rm Bondi}=L$, where $\dot M_{\rm Bondi}$ denotes the black-hole mass growth supplied by the SIDM Bondi channel, generic rotating SIDM inflow can spin up an initially nonrotating stellar-mass seed while it grows. The resulting spin evolution depends on the angular momentum carried by the inflow and on the relative contributions of SIDM and baryonic accretion. Consequently, the same dark-sector parameters that determine whether rapid mass growth occurs can also influence the final spin distribution of the supermassive population.  Our illustrative secular evolutions and conditional population maps show how different angular-momentum histories shift both the mass growth and the final Kerr configuration.  The special $L=J/M$ family is dynamically distinctive: the ratio $J/M$ remains constant, and therefore the dimensionless spin $\chi=J/M^2$ decreases as the mass increases.

The paper is organized as follows.  Section~\ref{sec:outline} summarizes the stationary Kerr flow equations and the mass and angular-momentum fluxes.  Section~\ref{sec:generic} develops the generic critical SIDM solution, determines its physical existence domain and universal rate, and then discusses the analytically tractable $\ell=0$ and $\ell=\chi$ subfamilies.  Section~\ref{sec:evolution} studies the coupled secular evolution of mass and spin, incorporates the numerical benchmark and finite-halo prescription, and presents the Bondi-radius and population illustrations.  Section~\ref{sec:discussion} summarizes the implications for rapid SMBH assembly and the limitations of the quasi-spherical treatment.  Appendix~\ref{app:kerrflow} gives the detailed derivation of the stationary Kerr flow, while Appendix~\ref{app:critical} collects the critical expansions, quartic analysis, and analytic subfamily checks.

\section{Outline of Bondi flow on a Kerr black hole}\label{sec:outline}

\subsection{Stationary axisymmetric perfect fluid}

We work in geometrized units $G=c=1$ and introduce dimensionless Kerr variables from the outset,
\begin{equation}
 \tau\equiv\frac{t}{M},\qquad x\equiv\frac{r}{M},\qquad
 \chi\equiv\frac{J}{M^2},\qquad \ell\equiv\frac{L}{M}.
 \label{eq:dimensionless}
\end{equation}
Here $M$ is the Kerr black-hole mass parameter and $J$ is the black-hole angular momentum; thus $M$ is sometimes denoted $M_{\rm BH}$.  We use only $M$ throughout the paper to avoid introducing a second notation for the same black-hole mass.  In the stationary problem, $M$ is a fixed background parameter, while in Sec.~\ref{sec:evolution} we promote it adiabatically to the slowly varying mass $M(t)$.  The quantity $L$ is the dimensional angular momentum carried by the fluid per unit black-hole energy accreted.  We also scale the line element as $\dd\hat s^2\equiv\dd s^2/M^2$.  In Boyer--Lindquist coordinates $(\tau,x,\theta,\phi)$ the Kerr metric is then
\begin{align}
 \dd\hat s^2={}&-\left(1-\frac{2x}{\Sigma}\right)\dd\tau^2
 -\frac{4\chi x\sin^2\theta}{\Sigma}\dd\tau\,\dd\phi
 +\frac{\Sigma}{\Delta}\dd x^2+\Sigma\dd\theta^2 \\
 &+\left(x^2+\chi^2+\frac{2\chi^2x\sin^2\theta}{\Sigma}\right)
 \sin^2\theta\,\dd\phi^2,
 \label{eq:kerr-dimless}
\end{align}
where, throughout the paper,
\begin{equation}
 \Sigma(x,\theta)\equiv x^2+\chi^2\cos^2\theta,\qquad
 \Delta(x)\equiv x^2-2x+\chi^2.
 \label{eq:SigmaDelta}
\end{equation}
The dimensionless horizon radii are
\begin{equation}
 x_\pm=1\pm\sqrt{1-\chi^2}.
 \label{eq:horizons-main}
\end{equation}
This convention is used in both the main text and the appendices.

The perfect-fluid stress tensor is
\begin{equation}
 T^{\mu\nu}=(\rho+p)u^\mu u^\nu+pg^{\mu\nu}
 =\rho_0 h u^\mu u^\nu+pg^{\mu\nu},
 \qquad h=\frac{\rho+p}{\rho_0}.
\end{equation}
Here $\rho_0$ is the local rest-mass density, whereas $\rho$ is the total energy density in units with $c=1$. Besides, $p$ is the pressure and $h$ is the enthalpy.  Correspondingly, the integrated quantity $M_{\rm rest}$ introduced below tracks the rest mass transported by the fluid and is distinct from the Kerr mass $M$. For an adiabatic, stationary, axisymmetric flow, the Euler equation yields a conserved energy integral and an axial integral.  In the dimensionless coordinates of Eq.~\eqref{eq:dimensionless} we write
\begin{equation}
 E\equiv-hu_\tau,\qquad
 \ell\equiv\frac{h u_\phi}{E}=-\frac{u_\phi}{u_\tau},\qquad
 L=M\ell.
 \label{eq:ELmain}
\end{equation}
Thus $L$ has dimensions of length in geometrized units and is the angular momentum delivered per unit black-hole mass-energy accreted.  We also use the dimensionless angular velocity
$\Omega\equiv\dd\phi/\dd\tau=M\,\dd\phi/\dd t=u^\phi/u^\tau$, for which
\begin{equation}
 \Omega
 =-\frac{g_{\tau\phi}+\ell g_{\tau\tau}}
 {g_{\phi\phi}+\ell g_{\tau\phi}}.
 \label{eq:omega-main}
\end{equation}

We adopt an equatorial quasi-spherical reduction with $\theta=\pi/2$ and $u^\theta=0$. The equatorial continuity equation gives a conserved radial flux proportional to $x^2\rho_0u_{\rm in}$. To define an effective total accretion rate, we extend this equatorial flux over a solid angle of $4\pi$. This normalization is a modeling assumption rather than an angular integration of a fully multidimensional Kerr flow. The coordinate component of the four-velocity satisfies $u^x=\dd x/\dd\hat\lambda<0$ for an accreting trajectory, where $\hat\lambda=\lambda/M$ is a dimensionless streamline parameter.  To use an inward-positive convention throughout, we define the radial inflow speed
\begin{equation*}
 u_{\rm in}\equiv-u^x=-\frac{\dd x}{\dd\hat\lambda}>0.
\end{equation*}
The continuity equation then gives
\begin{equation}
 x^2\rho_0u_{\rm in}={\rm const},\qquad
 \dot M_{\rm rest}=4\pi M^2x^2\rho_0u_{\rm in}.
 \label{eq:restmassflux}
\end{equation}
Here $M_{\rm rest}$ is a bookkeeping variable for the accumulated fluid rest mass crossing a constant-$x$ surface, and $\dot M_{\rm rest}>0$ is its inward flux.  It is important not to identify $M_{\rm rest}$ with the black-hole mass $M$.  Rest-mass conservation makes $\dot M_{\rm rest}$ independent of radius for a stationary flow, whereas the black-hole mass increase is determined by the \emph{energy} carried by that rest mass.  The distinction is analogous to the difference between baryon mass and gravitational mass in relativistic fluid dynamics.

\subsection{Energy and angular-momentum accretion rates}

The Killing currents associated with stationarity and axisymmetry are
\begin{equation}
 j_E^\mu=-T^\mu{}_{\tau},\qquad
 j_J^\mu=M\,T^\mu{}_{\phi},
\end{equation}
where the factor of $M$ restores the dimensional angular momentum associated with $L=M\ell$.  Since the signed coordinate component is $u^x=-u_{\rm in}$, the radial stress-tensor components are
\begin{equation}
 T^x{}_{\tau}=E\rho_0u_{\rm in},
 \qquad
 M T^x{}_{\phi}=-EL\rho_0u_{\rm in}.
\end{equation}
The energy current points toward decreasing $x$, while the angular-momentum current has the sign of $L$.  Defining the flux through a constant-$x$ surface as positive when it is directed inward, we denote the black-hole mass-energy growth supplied by the SIDM Bondi inflow by $\dot M_{\rm Bondi}$.  We use this notation to distinguish the Bondi channel from the phenomenological Eddington channel introduced below.  The corresponding fluxes are
\begin{equation}
 \dot M_{\rm Bondi}\equiv(\dot M)_{\rm Bondi}=E\dot M_{\rm rest},\qquad
 \dot J=EL\dot M_{\rm rest}=L\dot M_{\rm Bondi}.
 \label{eq:energyJflux}
\end{equation}
The first equality is the key distinction between rest mass and black-hole mass-energy.  Here $\dot M_{\rm Bondi}$ denotes the contribution to the growth rate of the same Kerr mass $M$ from Bondi accretion of SIDM; it is not a third mass variable.  The Bernoulli constant $E=-h u_\tau$ is the dimensionless Killing energy carried by the fluid per unit rest mass.  Hence, when considering the Bondi contribution, an element $\dd M_{\rm rest}$ of accreted rest mass changes the Kerr mass by
\[
 \dd M=E\,\dd M_{\rm rest}\qquad({\rm Bondi\ contribution}).
\]
Thus $M$ and $M_{\rm rest}$ have the same mass dimension; $E$ is dimensionless and accounts for the kinetic, internal, and gravitational binding contributions to the energy delivered to the hole.  If ordinary units are restored, the same statement is $\dd(Mc^2)=E\,c^2\dd M_{\rm rest}$ for the Bondi contribution.

The specific energy $E$ approaches unity in the cold critical limit, which is precisely the regime considered here for super-Eddington growth driven by SIDM accretion.  In this limit the Bondi-induced increment obeys $\dd M\simeq\dd M_{\rm rest}$, or equivalently $\dot M_{\rm Bondi}\simeq\dot M_{\rm rest}$.  Adding a phenomenological Eddington-limited baryonic contribution $\dot M_{\rm Edd}\equiv\Gamma M$, the total black-hole mass-growth rate can therefore be written, to leading critical order, as
\begin{equation}
 \dot M=\dot M_{\rm Edd}+\dot M_{\rm Bondi}
 \simeq \Gamma M+4\pi M^2x^2\rho_0u_{\rm in}.
 \label{eq:total-growth}
\end{equation}
Here $\Gamma$ is the specific Eddington mass-growth rate, with dimensions of inverse time. We introduce the Eddington contribution phenomenologically to account for baryonic mass growth. In the illustrative mass–spin evolution considered below, we neglect the angular momentum transferred by this baryonic channel and attribute the black-hole angular-momentum growth entirely to the SIDM Bondi inflow. This simplifying assumption isolates the rotational effects of SIDM accretion and is not intended as a general description of baryonic disk accretion. We retain $\rho_0$ in Eq.~\eqref{eq:total-growth} because the conserved continuity flux is the rest-mass flux; no replacement by the total energy density $\rho$ is required.

Eliminating $\dd M_{\rm rest}$ between the Bondi energy and angular-momentum fluxes gives
\begin{equation}
 \left.\frac{\dd J}{\dd M}\right|_{\rm Bondi}=L.
 \label{eq:dJdM}
\end{equation}
This relation is kinematical once the stationary streamline integrals are defined.  It is the fluid analog of the familiar statement that the black-hole mass and spin respond to the energy and angular momentum delivered through the horizon; see also general flux/backreaction discussions such as Ref.~\cite{Kimura2021}.

The horizon regularity condition for future-directed matter is most transparently written in dimensionless form,
\begin{equation}
 E\left(1-\Omega_H\ell\right)>0,
 \qquad \Omega_H=\frac{\chi}{2x_+},
 \label{eq:horizon-energy}
\end{equation}
where $\Omega_H$ is the dimensionless horizon angular velocity $M\Omega_H^{\rm phys}$. For positive Killing energy $E$, inward passage through the future horizon requires $1-\Omega_H\ell>0$. This provides a necessary condition on the accreting fluid's angular momentum, in addition to the sonic-point regularity and physical admissibility conditions discussed below.

\subsection{Sonic point and Bernoulli relation}

We use the following radial-subscript convention throughout the main text and the appendices.  The symbol $c_{\rm s}$ denotes the \emph{local adiabatic sound speed}; here the roman ``${\rm s}$'' is part of the name ``sound speed'' and does not denote a radial location.  By contrast, a separate subscript $s$ on any quantity denotes evaluation at the sonic point $x=x_s$,
\begin{equation*}
 Q_s\equiv Q(x_s),\qquad c_{{\rm s},s}\equiv c_{\rm s}(x_s).
\end{equation*}
A subscript $c$ denotes the value on the \emph{critical} sonic branch, for example $c_{{\rm s},c}\equiv c_{{\rm s},s}|_{\rm crit}$ and $x_c$ is the corresponding critical sonic coordinate.  Finally, a subscript $\infty$ denotes the asymptotic outer-halo state of the stationary problem, idealized by $x\to\infty$; thus $Q_\infty\equiv Q(x\to\infty)$.  We reserve this $\infty$ notation for the fluid boundary state and do not use it for late-time cosmological fixed points.

The local adiabatic sound speed is
\begin{equation}
 c_{\rm s}^2=\left(\frac{\partial p}{\partial\rho}\right)_\sigma
 =\frac{\rho_0}{h}\frac{\dd h}{\dd\rho_0},
\end{equation}
where $\sigma$ denotes the specific entropy held fixed in the adiabatic derivative.  The poloidal four-velocity norm, which first appears in the wind equation below, is defined already here.  In the equatorial quasi-spherical reduction it is
\begin{equation*}
 u_{\rm pol}^2=g_{xx}u_{\rm in}^2,
\end{equation*}
where $u_{\rm in}=-u^x>0$ is the inward-positive radial speed defined above.  Along the future-directed streamline, $\dd x/\dd\hat\lambda=-u_{\rm in}$.  The detailed derivation in Appendix~\ref{app:kerrflow} yields a wind equation
\begin{equation}
 \frac{\dd u_{\rm pol}}{\dd\hat\lambda}=\frac{\Ncal}{\Dcal},
\end{equation}
where $\Ncal$ and $\Dcal$ are respectively the numerator and denominator given explicitly in Eqs.~\eqref{eq:Napp} and \eqref{eq:Dapp}.  Regularity at the sonic point requires $\Ncal_s=\Dcal_s=0$.  Writing $x_s\equiv r_s/M$, $u_{{\rm in},s}\equiv u_{\rm in}(x_s)$, and $\Delta_s\equiv\Delta(x_s)$ with the dimensionless Kerr function $\Delta(x)=x^2-2x+\chi^2$ already defined in Eq.~\eqref{eq:SigmaDelta}. The denominator condition gives
\begin{equation}
 u_{{\rm in},s}^2=\frac{c_{{\rm s},s}^2\Delta_s}{x_s^2(1-c_{{\rm s},s}^2)}.
 \label{eq:sonicDmain}
\end{equation}
The numerator condition fixes $c_{{\rm s},s}^2$ as a function of the dimensionless sonic radius $x_s$, the black-hole spin $\chi$, and the inflow angular momentum $\ell$.  Define the radial functions
\begin{align}
 A(x;\chi,\ell)&=2(\chi-\ell)^2+x(\chi^2-\ell^2)+x^3,\label{eq:Adef}\\
 B(x;\chi)&=\chi^2+x(2x-3),\label{eq:Bdef}\\
 n(x;\chi,\ell)&=(\chi-\ell)^2(\chi^2-4x)
 +2x^2(\chi-2\ell)(\chi-\ell)-\ell^2x^3+x^4,
 \label{eq:ndef}
\end{align}
and their sonic-point values are denoted
$A_s\equiv A(x_s)$, $B_s\equiv B(x_s)$, $n_s\equiv n(x_s)$.  The sonic regularity condition then reads
\begin{equation}
 c_{{\rm s},s}^2=\frac{n_s}{A_sB_s}.
 \label{eq:css-generic-main}
\end{equation}

We denote the squared asymptotic sound speed by a single symbol throughout the rest of the paper,
\begin{equation}
 \zeta\equiv \left.c_{\rm s}^2\right|_{\infty}
 =\left[c_{\rm s}(x\to\infty)\right]^2.
 \label{eq:zeta-def}
\end{equation}
After this definition, we use $\zeta$ exclusively, rather than carrying a second asymptotic sound-speed symbol.  The small parameter $\zeta$ therefore measures the coldness of the outer halo rather than the sound speed at the sonic point: $\zeta=c_{{\rm s},\infty}^2$, whereas $c_{{\rm s},s}^2=c_{\rm s}^2(x_s)$.  In the cold critical regime $\zeta\ll1$, the Bernoulli constant satisfies\footnote{At the asymptotic boundary, $\mathcal G_\infty\to-1$ and $u_{{\rm pol},\infty}\to0$, so the Bernoulli equation gives $E=h_\infty$.  For the $\phi^4$ SIDM equation of state, $h_\infty=\sqrt{(1-\zeta)/(1-3\zeta)}=1+\zeta+O(\zeta^2)$, and hence $E=1+O(\zeta)$.}
\[
 E=1+O(\zeta),\qquad
 \dot M_{\rm Bondi}=\dot M_{\rm rest}\,[1+O(\zeta)],
\]
which makes precise the leading-order identification of the Bondi-induced black-hole mass growth with the rest-mass influx mentioned above.  The Bernoulli relation specialized to the $\phi^4$ fluid below gives
\begin{equation}
 \zeta=\frac{\Mcal_s-1}{\Mcal_s-3},\qquad
 \Mcal_s\equiv\frac{A_s}{x_s\Delta_s}
 (1-3c_{{\rm s},s}^2).
 \label{eq:cinf-generic-main}
\end{equation}
The symbol $\Mcal_s$ is \emph{not} the fluid Mach number.  The relativistic poloidal Mach number is instead defined from the local poloidal three-speed by
\begin{equation*}
 v_{\rm pol}^2\equiv\frac{u_{\rm pol}^2}{1+u_{\rm pol}^2},
 \qquad
 {\rm Ma}^2\equiv\frac{v_{\rm pol}^2}{c_{\rm s}^2}
 =\frac{u_{\rm pol}^2}{c_{\rm s}^2(1+u_{\rm pol}^2)}.
\end{equation*}
The denominator sonic condition implies ${\rm Ma}_s=1$.  Appendix~\ref{app:thermo-bern} derives Eq.~\eqref{eq:cinf-generic-main} step by step and again states the sonic and asymptotic conventions before using them.  Equations~\eqref{eq:css-generic-main} and \eqref{eq:cinf-generic-main} are the starting point for the critical analysis.

\section{Generic critical SIDM Bondi flow in Kerr}\label{sec:generic}

\subsection{$\phi^4$ SIDM and the outer boundary}\label{sec:sidm-eos}

The microscopic model is the same repulsively self-interacting real scalar considered in our previous work.  With the $(-,+,+,+)$ signature its Lagrangian density is
\begin{equation}
 \mathcal L_\phi
 =-\frac12\nabla_\mu\phi\nabla^\mu\phi
 -\frac12m^2\phi^2-\frac{\lambda_4}{4!}\phi^4,
 \label{eq:sidm-lagrangian}
\end{equation}
where $m$ is the scalar mass and $\lambda_4>0$ is the repulsive quartic coupling.  In the isotropic self-gravitating regime relevant for the fluid treatment, this theory gives the effective equation of state used in our previous work and Ref.~\cite{Feng2022},
\begin{equation}
 \frac{\rho}{\rhob}=3\frac{p}{\rhob}+4\sqrt{\frac{p}{\rhob}},
 \qquad \rhob=\frac{3m^4}{2\lambda_4}
 \quad (\hbar=c=1).
 \label{eq:eos-main}
\end{equation}
In terms of $c_{\rm s}^2$,
\begin{equation}
 \rho=\frac{4\rhob(2-3c_{\rm s}^2)c_{\rm s}^2}{(1-3c_{\rm s}^2)^2},
 \qquad
 p=\frac{4\rhob c_{\rm s}^4}{(1-3c_{\rm s}^2)^2}.
 \label{eq:rhop-main}
\end{equation}
At the outer boundary, the subscript $\infty$ has the meaning fixed above; for example $\rho_{0,\infty}\equiv\rho_0(x\to\infty)$.  Using the variable $\zeta$ defined in Eq.~\eqref{eq:zeta-def}, the adiabatic density relation is
\begin{equation}
 \frac{\rho_0}{\rho_{0,\infty}}
 =\frac{c_{\rm s}^2}{\zeta}
 \sqrt{\frac{(1-c_{\rm s}^2)(1-3\zeta)^3}
 {(1-\zeta)(1-3c_{\rm s}^2)^3}}.
 \label{eq:rho0ratio-main}
\end{equation}
For $\zeta\ll1$,
\begin{equation}
 \rho_\infty=8\rhob \zeta+O(\zeta^2),\qquad
 p_\infty=4\rhob \zeta^2+O(\zeta^3),
 \label{eq:rhoinf-smallz}
\end{equation}
so the finite-halo asymptotic state is effectively pressureless and $\rho_{0,\infty}\simeq\rho_\infty\simeq\rhodm$.  We derive the expansion, including the next coefficients and the corresponding expansion of $\rho_{0,\infty}$, in Appendix~\ref{app:thermo-bern}.

Following our previous work \cite{ChenLin2026}, we adopt the astrophysically motivated SIDM self-scattering range $0.1\lesssim \sigma_{\rm DM}/m\lesssim 1,{\rm cm^2,g^{-1}}$, which constrains the quartic coupling $\lambda$ and gives approximately
\begin{equation}
 \rhob\simeq 6.96\times10^{18}
 \left(\frac{m}{\mathrm{GeV}}\right)^{5/2}
 \mathrm{kg\,m^{-3}},
 \label{eq:rhoBmass}
\end{equation}
or equivalently $\rhob\simeq2.20\times10^{-4}(m/\mathrm{eV})^{5/2}\,\mathrm{kg\,m^{-3}}$.  As in Ref.~\cite{ChenLin2026}, this numerical normalization corresponds to adopting the fiducial coupling $\lambda_4\simeq50(m/{\rm GeV})^{3/2}$, chosen within the astrophysically motivated self-scattering range. 

\subsection{Generic critical condition}\label{sec:generic-condition}

The generic critical condition follows directly from Eq.~\eqref{eq:cinf-generic-main}.  We denote the critical member of the sonic family by $x_s=x_c$ and use a subscript $c$ for all quantities evaluated there, e.g. $A_c\equiv A(x_c)$, $n_c\equiv n(x_c)$, $\Delta_c\equiv\Delta(x_c)$, and $\Mcal_c\equiv\Mcal_s|_{x_s=x_c}$.  A critical flow has
\begin{equation}
 \zeta\to0
 \qquad\Longleftrightarrow\qquad
 \Mcal_c=1,
 \label{eq:Mcal1}
\end{equation}
provided $\Mcal_c-3$ remains finite.  Substituting the sonic relation~\eqref{eq:css-generic-main} into Eq.~\eqref{eq:Mcal1} gives a polynomial for the sonic coordinate $x_s$,
\begin{align}
 \mathscr C(x_s;\chi,\ell)={}&x_s^4+(\ell^2-6)x_s^3
 +(10\chi\ell-5\ell^2)x_s^2\nonumber\\
 &+\left[6(\chi-\ell)^2-\chi^2\ell^2\right]x_s
 -\chi^2(\chi-\ell)^2=0.
 \label{eq:criticalquartic}
\end{align}
The physical critical sonic radius $x_c$ is the appropriate outer root of this equation.  The algebra leading from $\Mcal_s=1$ to the polynomial is shown explicitly in Appendix~\ref{app:quartic-domain}.  Not every algebraic root is a physical sonic point.  The outer branch must additionally satisfy
\begin{equation}
 x_c>x_+=1+\sqrt{1-\chi^2},
 \qquad 0<c_{{\rm s},c}^2<\frac13,
 \qquad A_c>0,
 \label{eq:physical-root}
\end{equation}
plus the future-horizon condition~\eqref{eq:horizon-energy}.  Here $x_+\equiv r_+/M$ is the dimensionless outer Kerr-horizon radius.  These conditions place the sonic point outside the horizon, select the physical branch of the $\phi^4$ fluid, keep the Bernoulli factor real, and require positive energy flux through the future horizon.

\subsection{Existence domain and marginally bound angular momenta}

The physical interval can be understood analytically at its endpoints.  As an endpoint is approached, the outer critical branch has $c_{{\rm s},c}^2\to0$.  Combining this with $\Mcal_c=1$ gives $A_c=x_c\Delta_c$, while Eq.~\eqref{eq:css-generic-main} requires $n_c=0$.  Eliminating $x_c$ between these two endpoint equations yields candidate angular momenta.  Applying the outer-horizon condition and continuously connecting to the Schwarzschild interval $-4<\ell<4$ selects precisely the prograde and retrograde marginally bound values.  The complete endpoint algebra, including the discarded inner/horizon candidates, is given in Appendix~\ref{app:domain-derivation}.  Numerical continuation of the physical outer root between the analytic endpoints gives
\begin{equation}
 \ell_-(\chi)<\ell<\ell_+(\chi),
 \label{eq:ell-domain}
\end{equation}
where
\begin{equation}
 \ell_+(\chi)=2\left(1+\sqrt{1-\chi}\right),\qquad
 \ell_-(\chi)=-2\left(1+\sqrt{1+\chi}\right).
 \label{eq:ellpm}
\end{equation}
These are precisely the prograde and retrograde dimensionless energy-normalized specific angular momenta $\ell=L/M$ of marginally bound Kerr geodesics \cite{Bardeen1972}.  Their radii are
\begin{equation}
 x_{\rm mb}^{+}=2-\chi+2\sqrt{1-\chi},\qquad
 x_{\rm mb}^{-}=2+\chi+2\sqrt{1+\chi}.
 \label{eq:rmb}
\end{equation}
Substitution shows
\begin{equation}
 \mathscr C(x_{\rm mb}^{\pm};\chi,\ell_\pm)=0,
 \qquad n(x_{\rm mb}^{\pm};\chi,\ell_\pm)=0.
\end{equation}
Thus the fluid critical point approaches a zero-sound-speed marginally bound orbit at the edge of the physical domain.  The interpretation is transparent: once the fluid angular momentum exceeds the marginally bound capture threshold, a quasi-spherical inviscid inflow cannot connect smoothly to the horizon without additional angular-momentum transport, shocks, or multidimensional dynamics.  Those phenomena are familiar in the broader Kerr transonic literature \cite{Goswami2007,Das2012,ShaikhDas2018,DasNandi2019,Sukova2017}.

\subsection{Generic critical mass and angular-momentum fluxes}\label{sec:generic-rate}

The exact conserved rest-mass flux derived in Appendix~\ref{app:kerrflow} is
\begin{equation}
 \dot M_{\rm rest}=4\pi M^2\rho_{0,\infty}
 \frac{(1-\zeta)(c_{{\rm s},s}^2)^{3/2}}{\zeta}
 \frac{A_s^{3/2}}{x_s^{1/2}\Delta_s}.
 \label{eq:mdot-generic}
\end{equation}
Because Eq.~\eqref{eq:mdot-generic} is the central stationary-flow result, it is useful to sketch its origin here.  Evaluating continuity at the sonic point and using Eq.~\eqref{eq:sonicDmain} gives
\[
 \dot M_{\rm rest}=4\pi M^2x_s\rho_{0,s}
 \sqrt{\frac{c_{{\rm s},s}^2\Delta_s}{1-c_{{\rm s},s}^2}}.
\]
The density ratio~\eqref{eq:rho0ratio-main} expresses $\rho_{0,s}/\rho_{0,\infty}$ in terms of $c_{{\rm s},s}^2$ and $\zeta$.  Finally, the Bernoulli identity~\eqref{eq:cinf-generic-main} is equivalent to
\[
 \frac{x_s\Delta_s}{A_s}
 =\frac{(1-3c_{{\rm s},s}^2)(1-\zeta)}{1-3\zeta},
\]
which removes the remaining thermodynamic factors and yields Eq.~\eqref{eq:mdot-generic}.  Appendix~\ref{app:kerrflow} gives the complete derivation from the Euler equation.

\begin{figure*}[t]
\centering
\includegraphics[width=0.99\linewidth]{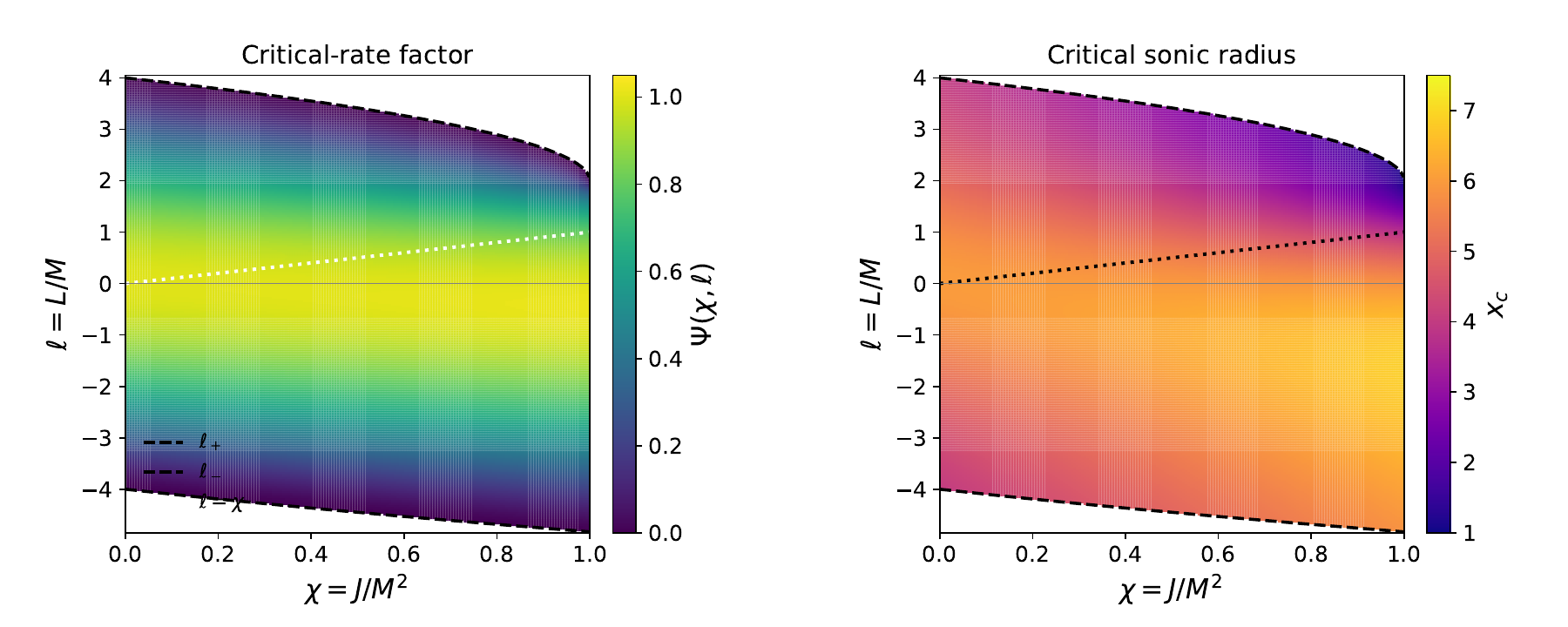}
\caption{Generic critical Kerr flows.  Left: the universal rate factor $\Psic(\chi,\ell)$ defined in Eq.~\eqref{eq:Psi-generic}.  Right: the critical sonic radius $x_c$, obtained as the physical outer root of the quartic equation~\eqref{eq:criticalquartic} subject to the physical conditions~\eqref{eq:physical-root}.  Dashed curves are the prograde and retrograde marginally bound angular-momentum boundaries~\eqref{eq:ellpm}; the dotted line is the analytic $\ell=\chi$ family discussed below.}
\label{fig:genericmap}
\end{figure*}

All quantities with a subscript $s$ in Eq.~\eqref{eq:mdot-generic} are evaluated at the ordinary sonic point of the stationary solution; only after the critical limit is taken do they become the corresponding $c$-subscripted quantities.  At a generic critical root, $\zeta\to0$ and Eq.~\eqref{eq:rhoinf-smallz} gives $\rho_{0,\infty}=8\rhob\zeta[1+O(\zeta)]$.  The limiting calculation is given in Appendix~\ref{app:critical-pole}.  The pole therefore cancels the asymptotic density and leaves
\begin{equation}
 \dot M_{{\rm rest},{\rm crit}}=64\pi\Psic(\chi,\ell)\rhob M^2
 \qquad (G=c=1),
 \label{eq:generic-rate}
\end{equation}
with
\begin{equation}
 \Psic(\chi,\ell)=\frac12
 \frac{(c_{{\rm s},c}^2)^{3/2}A_c^{3/2}}
 {x_c^{1/2}\Delta_c}.
 \label{eq:Psi-generic}
\end{equation}
For orientation, the quantities entering Eq.~\eqref{eq:Psi-generic} are fixed directly by the critical root $x_c$:
\[
 A_c=2(\chi-\ell)^2+x_c(\chi^2-\ell^2)+x_c^3,
 \qquad
 \Delta_c=x_c^2-2x_c+\chi^2,
\]
\[
 c_{{\rm s},c}^2=\frac{n_c}{A_cB_c},\qquad
 B_c=\chi^2+x_c(2x_c-3),\qquad n_c\equiv n(x_c;\chi,\ell),
\]
with $n(x;\chi,\ell)$ defined in Eq.~\eqref{eq:ndef}.
where $x_c$ is the physical outer root of Eq.~\eqref{eq:criticalquartic}.  Thus $\Psic$ is completely determined by $(\chi,\ell)$ once the physical critical root is selected.  Figure~\ref{fig:genericmap} summarizes the two central geometric outputs of the generic critical problem: the left panel shows $\Psic(\chi,\ell)$ from Eq.~\eqref{eq:Psi-generic}, while the right panel shows the corresponding $x_c$ obtained as the physical outer root of Eq.~\eqref{eq:criticalquartic} subject to the conditions in Eq.~\eqref{eq:physical-root}.

The density cancellation is therefore a property of the critical $\phi^4$ equation of state together with a regular Kerr sonic root; it is not tied to any special relation between the black-hole and fluid angular momenta.

In the pressureless critical limit $E=1+O(\zeta)$, so Eq.~\eqref{eq:energyJflux} gives $\dot M_{\rm Bondi}=\dot M_{\rm rest}[1+O(\zeta)]$.  The critical rest-mass rate can therefore be identified with the Bondi contribution to the black-hole mass-growth rate at leading order, but the two should not be confused away from this limit.  The angular-momentum rate follows simultaneously,
\begin{equation}
 \dot J_{\rm crit}
 =L\,\dot M_{{\rm Bondi},{\rm crit}}
 =64\pi\,\ell\,\Psic(\chi,\ell)\rhob M^3
 \qquad (G=c=1).
 \label{eq:Jdot-generic}
\end{equation}
Thus $(\chi,\ell)$ controls both the instantaneous growth efficiency and the direction in which the hole moves through mass--spin space.

\subsection{Microphysics--astrophysics matching and Kerr response}\label{sec:zeta-microphysics}

We can now combine the generic Kerr critical solution with the same SIDM microphysics used in our previous Schwarzschild analysis, while revising the normalization of the ambient outer-halo density.  The natural small parameter is the asymptotic squared sound speed $\zeta$ itself.  Setting $c_{\rm s}^2=\zeta$ in Eq.~\eqref{eq:rhop-main} gives the exact outer-boundary relation
\begin{equation}
 \frac{\rho_\infty}{\rhob}
 =\frac{4(2-3\zeta)\zeta}{(1-3\zeta)^2}.
 \label{eq:rhoinf-zeta-exact}
\end{equation}
For a finite halo, $p_\infty=O(\zeta^2)$ and $\rho_\infty=O(\zeta)$, so $\rho_\infty\simeq\rho_{0,\infty}\simeq\rhodm$.  Inverting the preceding equation gives
\begin{equation}
 \zeta=\frac13\left[1-
 \left(1+\frac{3\rhodm}{4\rhob}\right)^{-1/2}\right],
 \label{eq:zeta-rho-exact}
\end{equation}
and in the critical regime $\rhodm/\rhob\ll1$,
\begin{equation}
 \zeta\simeq\frac{1}{8}\frac{\rhodm}{\rhob}.
 \label{eq:zeta-rho-leading}
\end{equation}
The asymptotic density should be interpreted as the ambient SIDM density at the outer boundary of the Bondi flow, rather than as the density of the central SIDM core.  A transparent normalization is therefore the cosmological SIDM density multiplied by an environmental overdensity $\delta_{\rm env}$,
\begin{equation}
 \rhodm(z)\equiv\rho_\infty(z)
 =\delta_{\rm env}\,\Omega_{{\rm SIDM},0}\rho_{c0}(1+z)^3
 \simeq2.25\times10^{-27}\,\delta_{\rm env}(1+z)^3
 \;\mathrm{kg\,m^{-3}},
 \label{eq:rhoDM-fiducial}
\end{equation}
where we use $\Omega_{{\rm SIDM},0}\simeq0.264$ when the SIDM constitutes essentially all of the cosmological dark matter.  The factor $\delta_{\rm env}$ specifies the environment sampled by the Bondi outer boundary: $\delta_{\rm env}=1$ corresponds to the cosmic mean, while $\delta_{\rm env}\sim10^2$--$10^3$ represents a virialized or moderately enhanced outer-halo environment.  We adopt $\delta_{\rm env}=300$ as the fiducial benchmark below.

This normalization is substantially lower than the density used in our previous Schwarzschild study~\cite{ChenLin2026}, where $5\times10^{-22}(1+z)^3\Delta_{\rm vir}\,{\rm kg\,m^{-3}}$ was employed as an illustrative ambient value.  In retrospect, that choice should be regarded as an overly dense environmental benchmark rather than a virial-average ambient density.  Importantly, correcting this normalization does not alter the critical density cancellation or the environment-independent limiting Bondi rate; it changes the mapping between the outer criticality parameter and the microscopic scalar mass, and it changes the Bondi radius.

Combining Eq.~\eqref{eq:rhoDM-fiducial} with Eq.~\eqref{eq:rhoBmass} gives
\begin{equation}
 \zeta(z,m)\simeq1.28\times10^{-24}\,\delta_{\rm env}(1+z)^3
 \left(\frac{m}{\mathrm{eV}}\right)^{-5/2},
 \label{eq:zeta-m-redshift}
\end{equation}
or, normalized to our fiducial outer-halo environment,
\begin{equation}
 \zeta(z,m)\simeq3.84\times10^{-22}
 \left(\frac{\delta_{\rm env}}{300}\right)(1+z)^3
 \left(\frac{m}{\mathrm{eV}}\right)^{-5/2}.
 \label{eq:zeta-m-redshift-fid}
\end{equation}
Equivalently,
\begin{equation}
 \frac{m}{\mathrm{eV}}\simeq
 \left[\frac{3.84\times10^{-22}}{\zeta}
 \left(\frac{\delta_{\rm env}}{300}\right)(1+z)^3\right]^{2/5}.
 \label{eq:m-zeta-redshift}
\end{equation}
Thus, at fixed $\zeta$, the scalar mass inferred from the ambient matching scales as $m\propto\rho_\infty^{2/5}$.  Relative to the ambient normalization used in Ref.~\cite{ChenLin2026}, the same value of $\zeta$ corresponds, for $\delta_{\rm env}=300$, to a scalar mass smaller by a factor $\simeq7.3\times10^{-3}$.  This shift applies to the density-derived criticality bound; it should not be confused with the separate lower mass scale required for the Bondi term to dominate the cosmological growth equation.

These relations contain neither $\chi$ nor $\ell$: the particle physics and ambient outer state fix the same asymptotic $\zeta$ for every allowed Kerr critical branch.  Rotation enters only after the flow probes the Kerr sonic geometry.

Because $E=1+O(\zeta)$ on the critical branch, the leading critical rest-mass flux and Bondi-induced black-hole mass-growth rate share the same coefficient.  Thus the $\dot M_{\rm Bondi}$ notation introduced in Sec.~\ref{sec:outline} applies directly to the leading critical black-hole growth rate.  The corresponding Schwarzschild critical coefficient is
\begin{equation}
 K_0(m)\equiv64\pi\rhob(m)\frac{G^2}{c^3},
 \qquad
 \dot M_{\rm Bondi}^{\rm Schw}=K_0(m)M^2,
 \label{eq:K0def}
\end{equation}
with
\begin{equation}
 K_0(m)\simeq4.59\times10^{-10}
 \left(\frac{m}{\mathrm{eV}}\right)^{5/2}
 M_\odot^{-1}\,\mathrm{yr}^{-1}.
 \label{eq:Kspin}
\end{equation}
For generic Kerr flow it is therefore natural to define
\begin{equation}
 K_{\chi\ell}(m)\equiv\Psic(\chi,\ell)K_0(m),
 \qquad
 \dot M_{\rm Bondi}^{\rm Kerr}=K_{\chi\ell}(m)M^2.
 \label{eq:Kgeneric}
\end{equation}

The same outer value of $\zeta$ corresponds to a Kerr-dependent displacement of the ordinary sonic point from its critical root.  Let
\[
 y_s\equiv c_{{\rm s},s}^2,\qquad y_c\equiv c_{{\rm s},c}^2.
\]
For a simple critical root, Appendix~\ref{app:nearcritical} gives $\zeta=\Bcrit_{\chi\ell}(y_s-y_c)+O[(y_s-y_c)^2]$.  Inverting this relation and substituting the fiducial form Eq.~\eqref{eq:zeta-m-redshift-fid} gives
\begin{equation}
 y_s-y_c
 =\frac{\zeta}{\Bcrit_{\chi\ell}}+O(\zeta^2)
 \simeq
 \frac{3.84\times10^{-22}}{\Bcrit_{\chi\ell}}
 \left(\frac{\delta_{\rm env}}{300}\right)(1+z)^3
 \left(\frac{m}{\mathrm{eV}}\right)^{-5/2}.
 \label{eq:B-generic}
\end{equation}
Thus the direct $\zeta$--$m$ matching is angular-momentum independent, whereas $\Bcrit_{\chi\ell}$ determines how that same outer state is represented locally near the Kerr critical sonic point.

A compact explicit form of the response coefficient is
\begin{align}
 \Bcrit_{\chi\ell}
 ={}&\frac{A_c^2B_c}{2x_c\Delta_c}
 \frac{
 [ (\Delta_c+x_c\Delta'_c)B_c+x_c\Delta_c B'_c]
 -[A'_cB_c+A_cB'_c-3n'_c]
 }
 {n'_cA_cB_c-n_c(A'_cB_c+A_cB'_c)},
 \label{eq:B-explicit-main}
\end{align}
where a prime denotes differentiation with respect to $x$, followed by evaluation at the physical critical root $x=x_c$:
\begin{equation}
\begin{aligned}
 A'_c&=\chi^2-\ell^2+3x_c^2,\qquad
 B'_c=4x_c-3,\qquad
 \Delta'_c=2(x_c-1),\\
 n'_c&=-4(\chi-\ell)^2
 +4x_c(\chi-2\ell)(\chi-\ell)
 -3\ell^2x_c^2+4x_c^3.
\end{aligned}
\label{eq:derivs-B-main}
\end{equation}
At fixed local displacement $y_s-y_c$ and fixed halo density, Eq.~\eqref{eq:B-generic} also gives the useful comparison
\begin{equation}
 \frac{m_{\rm sonic}(\chi,\ell)}{m_{\rm sonic}(0,0)}
 =\left(\frac{\Bcrit_{00}}{\Bcrit_{\chi\ell}}\right)^{2/5},
 \label{eq:generic-mass-sonic}
\end{equation}
which is a local sonic-response diagnostic, not a modification of the universal relation~\eqref{eq:zeta-m-redshift}.

\begin{figure*}[t]
\centering
\includegraphics[width=0.99\linewidth]{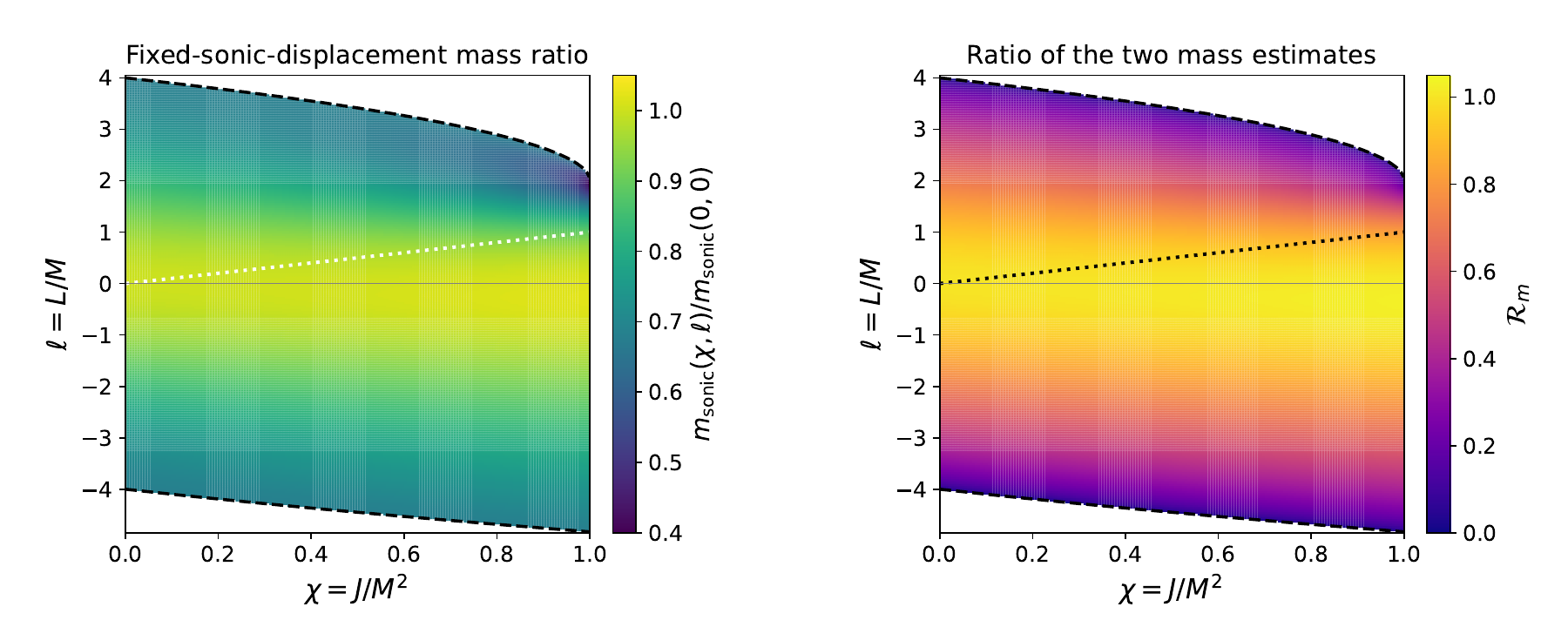}
\caption{Mass-scale diagnostics across the generic critical Kerr domain.  Left: the normalized scalar-mass ratio $m_{\rm sonic}(\chi,\ell)/m_{\rm sonic}(0,0)$ from Eq.~\eqref{eq:generic-mass-sonic}, which gives the mass shift required to hold the local sonic displacement $y_s-y_c$ fixed at fixed halo density.  Right: the ratio of the right-hand sides of Eqs.~\eqref{eq:generic-mass-sonic} and \eqref{eq:generic-mass-growth}, $\mathcal R_m=[\Bcrit_{00}\Psic(\chi,\ell)/\Bcrit_{\chi\ell}]^{2/5}$, comparing the fixed-sonic-displacement mass estimate with the fixed-growth estimate.  Values near unity indicate comparable Kerr corrections in the two criteria.  Dashed curves show the prograde and retrograde marginally bound angular-momentum boundaries, and the dotted line denotes the analytic $\ell=\chi$ family.  Both panels are restricted to the physical outer critical branch.}
\label{fig:generic-mass-ratios}
\end{figure*}

A different question is the scalar mass required to keep the Bondi growth coefficient fixed.  Since $K_{\chi\ell}\propto\Psic(\chi,\ell)m^{5/2}$,
\begin{equation}
 \frac{m_{\rm grow}(\chi,\ell)}{m_{\rm grow}(0,0)}
 =\Psic(\chi,\ell)^{-2/5}.
 \label{eq:generic-mass-growth}
\end{equation}
The two Kerr-dependent functions therefore have distinct roles: $\Psic(\chi,\ell)$ fixes the critical growth efficiency, whereas $\Bcrit_{\chi\ell}$ fixes the local sonic response to the angular-momentum-independent asymptotic parameter $\zeta$.  Figure~\ref{fig:generic-mass-ratios} compares the corresponding scalar-mass shifts over the full physical critical domain.  To compare the two criteria directly, define the ratio of the right-hand sides of Eqs.~\eqref{eq:generic-mass-sonic} and \eqref{eq:generic-mass-growth},
\[
 \mathcal R_m(\chi,\ell)
 \equiv
 \frac{m_{\rm sonic}(\chi,\ell)/m_{\rm sonic}(0,0)}
 {m_{\rm grow}(\chi,\ell)/m_{\rm grow}(0,0)}
 =\left[\frac{\Bcrit_{00}\Psic(\chi,\ell)}{\Bcrit_{\chi\ell}}\right]^{2/5}.
\]
Values $\mathcal R_m\simeq1$ indicate that the two criteria imply comparable Kerr corrections to the scalar mass, whereas $\mathcal R_m<1$ means that keeping the Bondi growth coefficient fixed requires the larger mass shift.

\subsection{Analytic subfamilies: $\ell=0$ and $L=J/M$}\label{sec:special}

The generic formulas contain two particularly useful analytic slices.  They provide nontrivial checks on numerical root selection and help separate genuine frame-dragging effects from the inflow's centrifugal effect.

\paragraph{Zero conserved fluid angular momentum.}
For $\ell=0$, Eq.~\eqref{eq:criticalquartic} factorizes,
\begin{equation}
 \mathscr C(x_s;\chi,0)
 =(x_s^2-\chi^2)(x_s^2-6x_s+\chi^2).
\end{equation}
The outer physical root is
\begin{equation}
 x_c=3+\sqrt{9-\chi^2},
 \qquad
 c_{{\rm s},c}^2=\frac{1}{x_c+3},
 \label{eq:ell0root}
\end{equation}
and substitution into Eq.~\eqref{eq:Psi-generic} gives the exact identity
\begin{equation}
 \Psic(\chi,0)=1.
 \label{eq:Psiell0}
\end{equation}
Frame dragging therefore shifts the critical sonic point but leaves the universal critical rate equal to the Schwarzschild value when the fluid carries no conserved angular momentum.  This is a useful warning against interpreting the rate as a function of black-hole spin alone.

\paragraph{Matched corotating analytic branch $L=J/M$.}
The second analytic family is
\begin{equation}
 L=\frac{J}{M}=\chi M\qquad\Longleftrightarrow\qquad \ell=\chi.
 \label{eq:specialbranch}
\end{equation}
Equation~\eqref{eq:css-generic-main} then reduces to
\begin{equation}
 c_{{\rm s},s}^2=\frac{x_s-\chi^2}{\chi^2+x_s(2x_s-3)},
 \label{eq:css-special}
\end{equation}
and the critical quartic factorizes along the sonic family as $\mathscr C(x_s;\chi,\chi)=-x_sP_\chi(x_s)$, where
\begin{equation}
 P_\chi(X)=X^2(6-X)-\chi^2X(X+5)+\chi^4.
 \label{eq:Pchi}
\end{equation}
The physical critical radius is the largest root
\begin{equation}
 P_\chi(x_c)=0,
 \label{eq:Pc0}
\end{equation}
which interpolates from $x_c=6$ in Schwarzschild to $x_c=2+\sqrt3$ at extremality.  The corresponding sonic sound speed rises from $1/9$ to $2/\sqrt3-1$.

On this line, the generic rate factor reduces to the closed form
\begin{equation}
 \Fspin(\chi)=
 \frac{x_c^2(x_c-\chi^2)^{3/2}}
 {4[x_c(x_c-3)+2\chi^2]\sqrt{\chi^2+x_c(2x_c-3)}},
 \label{eq:Fspin-main}
\end{equation}
so that the critical rest-mass flux is
\begin{equation}
 \dot M_{{\rm rest},{\rm crit}}
 =64\pi\Fspin(\chi)\rhob\frac{G^2M^2}{c^3}.
 \label{eq:critical-special-SI}
\end{equation}
The small-spin expansion is
\begin{equation}
 \Fspin(\chi)=1-\frac{\chi^2}{6}-\frac{11\chi^4}{432}+O(\chi^6),
 \label{eq:Fsmallspin}
\end{equation}
and $\Fspin(1)=0.7907157\ldots$.  Thus this particular corotating slice suppresses the rate by at most about $21\%$.  We collect the exact Bernoulli expression and precritical mass-flux formula for this family in Appendix~\ref{app:special-expansion}.

For comparison with the scalar-mass criteria, define
\begin{equation}
 K_\chi(m)\equiv\Fspin(\chi)K_0(m).
 \label{eq:Kchi-special}
\end{equation}
At a fixed desired growth coefficient,
\begin{equation}
 \frac{m_{\rm grow}(\chi)}{m_{\rm grow}(0)}
 =\Fspin(\chi)^{-2/5},
 \label{eq:mgrowth-ratio}
\end{equation}
which is $1.098$ at extremality.  Equivalently, on the same analytic line the local sonic-response coefficient may be written as
\begin{equation}
 \zeta=\Bcrit_\chi
 \left(c_{{\rm s},s}^2-c_{{\rm s},c}^2\right)
 +O\!\left[\left(c_{{\rm s},s}^2-c_{{\rm s},c}^2\right)^2\right],
 \label{eq:Bspecial}
\end{equation}
with $\Bcrit_0=9/16$ and $\Bcrit_1=3/4$.  This coefficient helps diagnose the local shape of the sonic branch, but it is not needed to infer the scalar mass from the halo boundary condition; the universal $\zeta$--$m$ relation in Eq.~\eqref{eq:zeta-m-redshift}.  The detailed special-line expansion is given in Appendix~\ref{app:special-expansion}.  Figure~\ref{fig:massspin} shows $\Fspin$, the fixed-growth scalar-mass shift, and the normalized response coefficient $\Bcrit_\chi/\Bcrit_0$; read it as one slice through the full two-parameter critical domain in Fig.~\ref{fig:genericmap}.

\begin{figure}[t]
\centering
\includegraphics[width=0.78\linewidth]{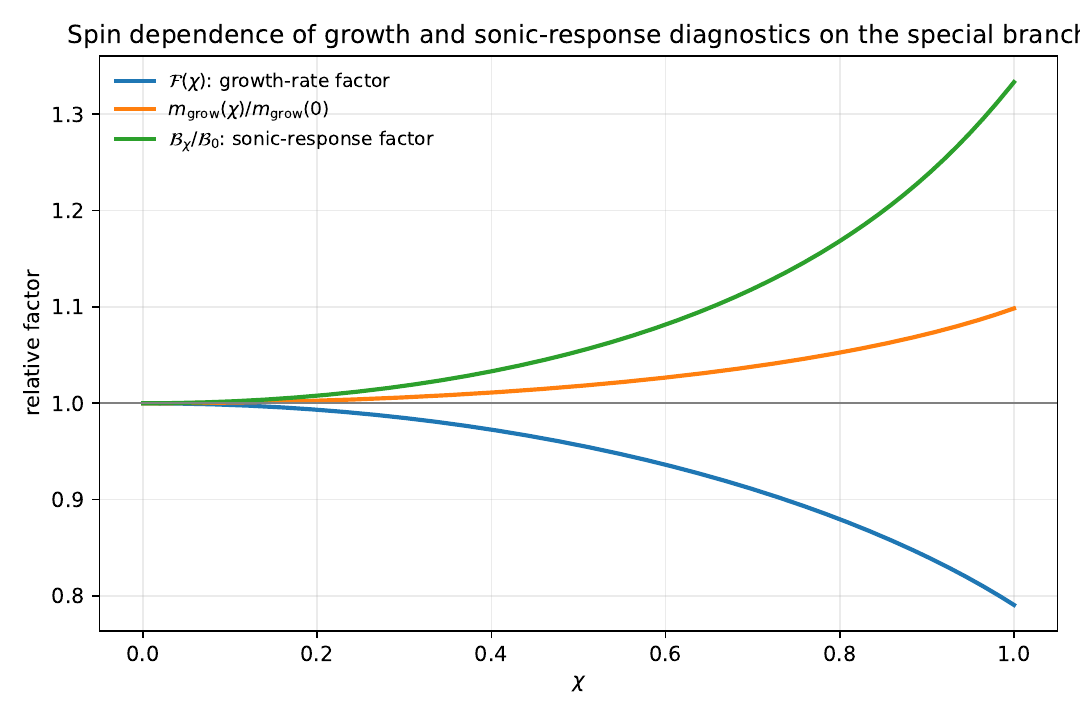}
\caption{Analytic diagnostics on the special $L=J/M$ ($\ell=\chi$) slice of the generic critical domain.  Shown are the critical-rate factor $\Fspin$, the scalar-mass shift required to keep the Bondi growth coefficient fixed, and the normalized local sonic-response coefficient $\Bcrit_\chi/\Bcrit_0$.  The last quantity characterizes the local approach to the critical sonic point and does not modify the angular-momentum-independent outer relation $\zeta(m)$ in Eq.~\eqref{eq:zeta-m-redshift}.  This figure is a one-dimensional benchmark; the generic Kerr problem is governed by $\Psic(\chi,\ell)$ in Fig.~\ref{fig:genericmap}.}
\label{fig:massspin}
\end{figure}

\section{Mass, angular momentum, and population evolution}\label{sec:evolution}

The stationary Bondi flux relation~\eqref{eq:dJdM} determines the angular momentum delivered by the dark-matter component, while the total black-hole mass may also contain the Eddington term in Eq.~\eqref{eq:total-growth}.  Throughout this section $M$ is the slowly evolving Kerr black-hole mass, not the accumulated fluid rest mass $M_{\rm rest}$.  Since $J=\chi M^2$, $L=\ell M$, and $\dot J=L\dot M_{\rm Bondi}$, differentiation gives the exact relation
\begin{equation}
 \frac{\dd\chi}{\dd\ln M}
 =\ell f_{\rm Bondi}-2\chi,
 \qquad
 f_{\rm Bondi}\equiv\frac{\dot M_{\rm Bondi}}{\dot M}.
 \label{eq:spin-master}
\end{equation}
The factor $f_{\rm Bondi}$ is the instantaneous fraction of the black-hole mass growth supplied by the SIDM Bondi channel.  In the super-Eddington regime of primary interest here, the critical Bondi term dominates, $f_{\rm Bondi}\simeq1$, and Eq.~\eqref{eq:spin-master} reduces to
\begin{equation}
 \frac{\dd\chi}{\dd\ln M}\simeq\ell-2\chi.
 \label{eq:spin-master-bondi}
\end{equation}
The $+\ell$ term is the dimensionless spin supplied by the accreting SIDM, whereas the $-2\chi$ term reflects the dilution of $J/M^2$ as the black hole mass increases.  In the Bondi-dominated limit and for $\dot M>0$, this is summarized compactly by
\[
 \operatorname{sgn}(\dot\chi)=\operatorname{sgn}(\ell-2\chi).
\]
Thus $\ell=2\chi$ is the instantaneous Bondi-dominated spin fixed point: the hole spins up for $\ell>2\chi$ and spins down for $\ell<2\chi$.  More generally, the fixed point is shifted by the factor $f_{\rm Bondi}$ in Eq.~\eqref{eq:spin-master}.  The final Kerr spin therefore depends on both the angular-momentum history of the SIDM inflow and the fraction of the total mass growth supplied by the Bondi channel.  The fixed-$\ell$ analytic benchmark below uses the super-Eddington approximation $f_{\rm Bondi}\simeq1$.

\subsection{Numerical benchmark, Bondi radius, and finite SIDM supply}\label{sec:numerical-benchmark}

The numerical curves below are intended to display the relative Kerr and angular-momentum effects with the minimum additional astrophysical modeling.  We adopt Planck-like cosmological parameters
\begin{equation}
 H_0=67.4\,\mathrm{km\,s^{-1}\,Mpc^{-1}},\qquad
 \Omega_m=0.315,\qquad \Omega_\Lambda=0.685,\qquad
 \Omega_r=9.2\times10^{-5}.
\end{equation}
We initialize the single-seed evolutions at $z_i=30$ with $M_i=10M_\odot$, adopting a fiducial SIDM particle mass $m=10^{-2}\,\mathrm{eV}$.  We retain this growth benchmark after revising the ambient density because the two constraints play different roles.  The revised density makes the asymptotic flow \emph{more} deeply critical at fixed $m$: for $\delta_{\rm env}=300$ and $m=10^{-2}\,\mathrm{eV}$, Eq.~\eqref{eq:zeta-m-redshift-fid} gives $\zeta\simeq1.1\times10^{-12}$ already at $z=30$.  Using the previous near-critical variable $\epsilon_{\rm old}$, for which $\zeta\simeq(9/16)\epsilon_{\rm old}$, the illustrative requirement $\epsilon_{\rm old}\lesssim10^{-3}$ over $7\le z\le30$ now implies only $m\gtrsim3\times10^{-6}\,\mathrm{eV}$.  By contrast, the onset of rapid Bondi growth is controlled by $K_0\propto m^{5/2}$ and still favors masses of order $10^{-2}\,\mathrm{eV}$ or larger.  The revised ambient normalization therefore relaxes the criticality bound without weakening the growth-motivated fiducial choice.

The Eddington growth coefficient is fixed at $\Gamma=4.5\times10^{-8}\,\mathrm{yr}^{-1}$. Throughout the numerical analysis, we use the critical accretion coefficient $K_0(m)$ derived in Eqs.~\eqref{eq:K0def} and ~\eqref{eq:Kspin}, together with the Kerr response factor $\Psi(\chi,\ell)$ defined in Eq.~\eqref{eq:Kgeneric}. For the fiducial particle mass, this gives $K_0=4.59\times10^{-15}\,M_\odot^{-1}\mathrm{yr}^{-1}$. We evaluate the Kerr response at the instantaneous $(\chi,\ell)$ values using the physical outer critical root. We introduce the finite-halo supply prescription below.

To model the finite dark-matter reservoir, we take $\Delta_{\rm vir}=300$ and define the halo virial radius by
$R_{\rm vir}(z)\equiv[3M_{\rm halo}(z)/(4\pi\Delta_{\rm vir}\rho_c(z))]^{1/3}$, where
$\rho_c(z)\equiv3H^2(z)/(8\pi G)$ is the cosmological critical density. We then use the same illustrative host prescription throughout this section,
\begin{equation}
 M_{\rm halo}(z)=5\times10^9
 \left(\frac{11}{1+z}\right)^3M_\odot,
 \qquad r_{\rm core}=0.05R_{\rm vir},
 \label{eq:hostmodel}
\end{equation}
so $r_{\rm core}$ is an illustrative core scale equal to five percent of the virial radius. Once the Bondi capture region grows to the scale of the inner core, finite-supply effects become relevant. We impose a cumulative SIDM mass budget, requiring the total mass accreted through the SIDM Bondi channel not to exceed $0.9M_{\rm halo}(z)$. This prescription is not part of the stationary Kerr solution; it provides a phenomenological constraint on the total dark-matter mass available for black-hole growth. 

The evolution of the Bondi radius is therefore essential to the SMBH-formation mechanism.  The critical $M^2$ law is a local accretion law embedded in a finite dark-matter halo, whereas $r_B$ measures the scale from which the black hole can gravitationally capture the ambient SIDM.  As the black hole grows, $r_B$ expands.  Once it becomes comparable to the dense inner halo, the assumption of an effectively unlimited, homogeneous reservoir breaks down, and the available SIDM supply, rather than the local critical rate, controls subsequent growth.  Tracking $r_B(z)$ therefore tells us when the rapid Bondi-dominated phase reaches the finite halo core and becomes supply-limited.

We use the same Bondi-radius convention as in our previous work.  At the outer boundary the physical sound speed is
$a_\infty=c\,c_{{\rm s},\infty}=c\sqrt{\zeta}$.  A simple way to estimate the capture scale is to equate the asymptotic sound speed to the Newtonian escape speed at $r_B$,
\begin{equation*}
 a_\infty^2\simeq v_{\rm esc}^2(r_B)=\frac{2GM}{r_B}.
\end{equation*}
Since $a_\infty^2=c^2\zeta$, this immediately gives, at leading critical order,
\begin{equation}
 r_B=\frac{2GM}{c^2\zeta}.
 \label{eq:rBdef}
\end{equation}
Using the finite-halo matching $\zeta\simeq\rhodm/(8\rhob)$ from Eq.~\eqref{eq:zeta-rho-leading} then yields the angular-momentum-independent leading relation
\begin{equation}
 r_B\simeq16\frac{\rhob}{\rhodm}\frac{GM}{c^2}.
 \label{eq:rBcritical}
\end{equation}
Thus the leading capture scale is fixed by the same outer-halo and SIDM microphysics that determine $\zeta$.  The Kerr parameters affect $r_B(z)$ indirectly through the mass history $M(z)$, while subleading near-critical corrections can depend on $\Bcrit_{\chi\ell}$.

\begin{figure}[htbp]
\centering
\includegraphics[width=0.80\linewidth]
{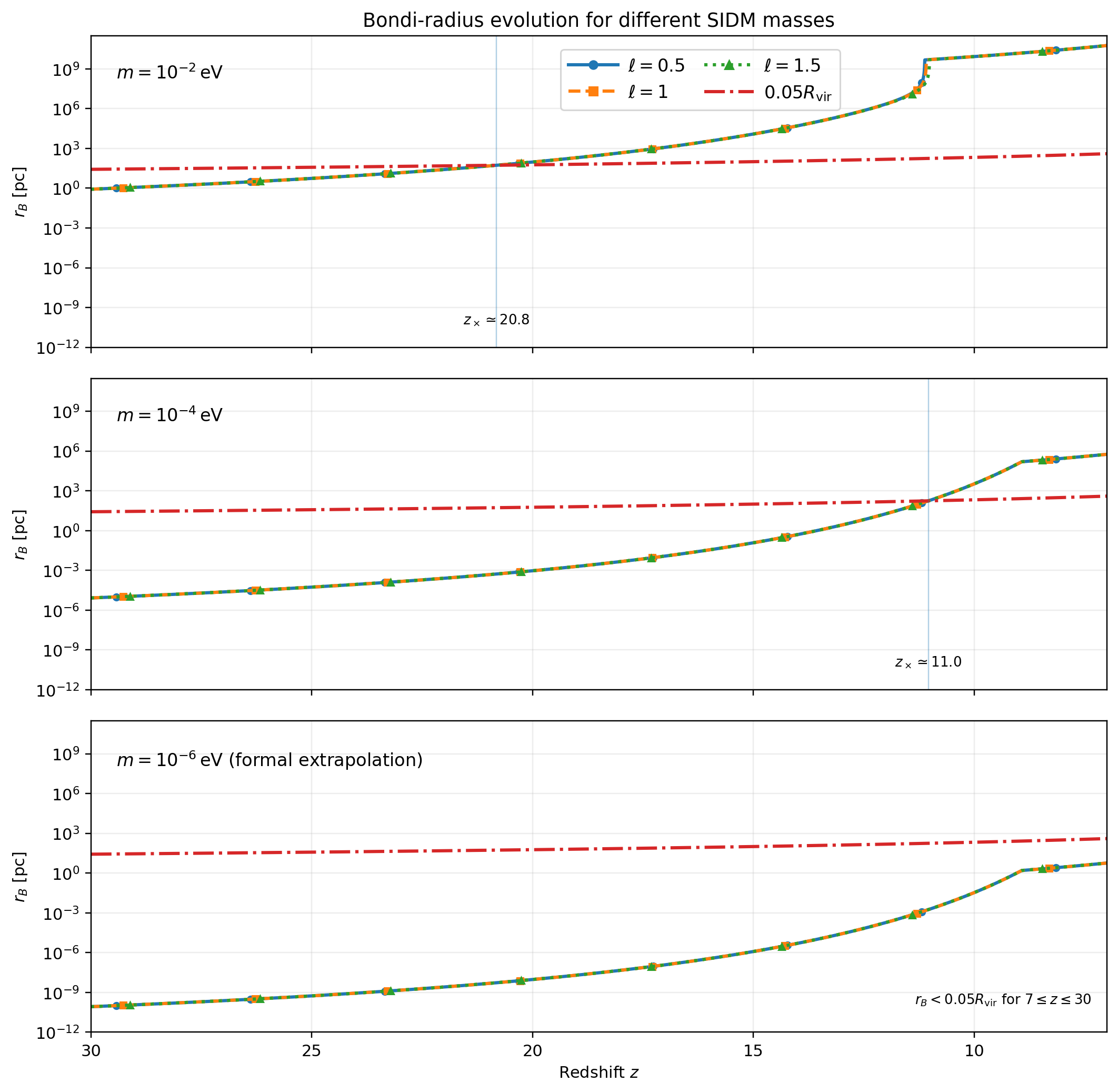}
\caption{Bondi-radius evolution using the revised ambient-density normalization with $\delta_{\rm env}=300$, for $m=10^{-2},10^{-4},10^{-6}\,\mathrm{eV}$ (top to bottom).  In each panel the three curves correspond to $\ell=0.5,1,1.5$, with distinct line styles and markers; the dash-dotted curve is the illustrative core scale $r_{\rm core}=0.05R_{\rm vir}$.  The mass history is recomputed self-consistently for each scalar mass using $K_0(m)\propto m^{5/2}$.  For $m=10^{-2}\,\mathrm{eV}$ the Bondi radius reaches the core scale near $z\simeq20.8$.  For $m=10^{-4}\,\mathrm{eV}$ the crossing is delayed to $z\simeq11.0$, whereas for $m=10^{-6}\,\mathrm{eV}$ one has $r_B<r_{\rm core}$ throughout $7\le z\le30$.  The three $\ell$ curves remain almost coincident except near the rapid-growth transition because the angular-momentum dependence enters mainly through the small differences in the mass histories.  The lowest panel is a formal extrapolation of the radius scaling below the approximate criticality bound and should not be interpreted as a fully self-consistent critical-flow benchmark.}
\label{fig:genericrb}
\end{figure}

Using the revised ambient density in Eq.~\eqref{eq:rhoDM-fiducial} together with Eq.~\eqref{eq:rBcritical}, we obtain
\begin{equation}
r_B(z,m)\simeq2.50\times10^8
 \left(\frac{\delta_{\rm env}}{300}\right)^{-1}
 \left(\frac{m}{\mathrm{eV}}\right)^{5/2}
 \frac{M(z)}{M_\odot}(1+z)^{-3}\,\mathrm{pc}.
 \label{eq:rBbenchmark}
\end{equation}
For the fiducial $m=10^{-2}\,\mathrm{eV}$ this is $r_B\simeq2.50\times10^3[M(z)/M_\odot](1+z)^{-3}\,$pc.  The much larger radius compared with our previous estimate is the most important numerical consequence of the revised ambient normalization.  Figure~\ref{fig:genericrb} shows Eq.~\eqref{eq:rBbenchmark} for three representative scalar masses, $m=10^{-2},10^{-4},10^{-6}\,\mathrm{eV}$, and for $\ell=0.5,1,1.5$.  For each value of $m$ the secular mass history is recomputed with the corresponding critical growth coefficient $K_0(m)\propto m^{5/2}$, rather than merely rescaling $r_B$ at fixed $M(z)$.  The crossing $r_B\simeq r_{\rm core}$ should be interpreted as the point at which the homogeneous ambient-boundary approximation first becomes sensitive to the finite inner halo, rather than as an instantaneous exhaustion of the SIDM reservoir.  The secular evolution below continues to implement finite supply through the cumulative halo-mass cap.  The $m=10^{-6}\,\mathrm{eV}$ panel is shown only as a formal extrapolation of the radius scaling, since this mass lies below the approximate criticality bound inferred above for the fiducial environmental overdensity.

\clearpage

\subsection{Fixed-$\ell$ benchmark: spin-up from nonrotating seeds}

The stationary analysis in Sec.~\ref{sec:generic} determines the critical accretion coefficient for each instantaneous pair $(\chi,\ell)$, but it does not prescribe how the surrounding SIDM's angular momentum changes during cosmological growth.  This subsection therefore introduces a simple benchmark angular-momentum history that connects the stationary critical-flow solutions to an explicit secular evolution.  We assume that the dimensionless inflow angular momentum $\ell=L/M$ is approximately constant during a growth episode.  This assumption is not meant to be a unique astrophysical prediction; it is the simplest choice that isolates how a nonzero inflow angular momentum both spins up the hole and changes the critical Bondi coefficient $\Psic(\chi,\ell)$.

For the analytic benchmark we now set $f_{\rm Bondi}=1$, appropriate to the Bondi-dominated super-Eddington regime.  With fixed $\ell$, Eq.~\eqref{eq:spin-master-bondi} is then equivalent to
\begin{equation}
 \frac{\dd J}{\dd M}=\ell M.
 \label{eq:dJdM-fixedell}
\end{equation}
Let the growth episode begin with black-hole mass $M_i$ and spin $\chi_i$, so that $J_i=\chi_i M_i^2$.  Integrating Eq.~\eqref{eq:dJdM-fixedell} gives
\[
 J(M)=J_i+\frac{\ell}{2}(M^2-M_i^2),
\]
and therefore
\[
 \chi(M)=\frac{\ell}{2}+
 \left(\chi_i-\frac{\ell}{2}\right)
 \left(\frac{M_i}{M}\right)^2.
\]
For the initially Schwarzschild seeds used in our benchmark, $\chi_i=0$ and $J_i=0$.  The previous expressions reduce to
\begin{equation}
 J(M)=\frac{\ell}{2}(M^2-M_i^2),\qquad
 \chi(M)=\frac{\ell}{2}\left[1-\left(\frac{M_i}{M}\right)^2\right].
 \label{eq:spinup-fixedell}
\end{equation}
Here $M_i$ is the seed mass at the beginning of the growth episode; in the numerical examples of Sec.~\ref{sec:numerical-benchmark}, $M_i=10M_\odot$ at $z_i=30$.  Equation~\eqref{eq:spinup-fixedell} gives $J$ and $\chi$ at \emph{any} intermediate mass $M$, not only their final values.  If the episode ends at $M=M_f$, then the final quantities are simply $J_f=J(M_f)$ and $\chi_f=\chi(M_f)$.  For sufficiently large growth,
\begin{equation}
 \chi\longrightarrow\frac{\ell}{2}
 \qquad (M\gg M_i).
 \label{eq:chi-asymptote}
\end{equation}
Thus, $\chi=\ell/2$ is the asymptotic fixed point in the Bondi-dominated limit. For $0<\ell<2$, this limiting spin is subextremal. Equation~\eqref{eq:spinup-fixedell} shows how rotating SIDM accretion can spin up an initially Schwarzschild stellar-mass PBH while rapidly increasing its mass. When the Eddington contribution is retained, however, the spin evolution must be determined from the full mass and angular-momentum balance. 

For the numerical benchmark, we retain both the Eddington and critical Kerr Bondi contributions to the black-hole mass growth. Using the coefficient $K_0(m)$ defined in Eq.~\eqref{eq:K0def}, we obtain
\begin{equation}
\begin{aligned}
 \dot M &= \Gamma M+K_0(m)\Psi(\chi,\ell)M^2,\\
\dot\chi &= (\ell-2\chi)K_0(m)\Psi(\chi,\ell)M-2\Gamma\chi.
\end{aligned}
\label{eq:growth-generic}
\end{equation}
The second equation follows from Eq.~\eqref{eq:spin-master}, assuming that the baryonic Eddington component contributes to the black-hole mass but carries no net angular-momentum torque. The two equations are integrated simultaneously, with the Kerr response factor $\Psi(\chi,\ell)$ evaluated at each instantaneous value of $(\chi,\ell)$. We apply the finite SIDM supply prescription introduced in Sec.~\ref{sec:numerical-benchmark} consistently to the SIDM contributions to both mass and angular-momentum growth.

In the Bondi-dominated regime, $f_{\rm Bondi}\simeq1$, the spin evolution approaches the analytic trajectory of Eq.~\eqref{eq:spinup-fixedell}, and the limiting spin is $\chi_{\rm fp}=\ell/2$. For the representative values $\ell=0.5,1,1.5$, these limiting spins are $(0.25,0.50,0.75)$, respectively. The corresponding critical Kerr response factors for three representative values of $\ell$ are summarized below.
\begin{center}
\begin{tabular}{cccc}
\toprule
$\ell$ & $\chi_{\rm fp}$ & $x_c$ & $\Psic(\chi_{\rm fp},\ell)$\\
\midrule
0.5 & 0.25 & 5.739 & 0.9682\\
1.0 & 0.50 & 4.949 & 0.8660\\
1.5 & 0.75 & 3.593 & 0.6614\\
\bottomrule
\end{tabular}
\end{center}
Larger co-rotating angular momentum increases the limiting spin while suppressing the critical mass-growth coefficient via the centrifugal contribution to $\Psi$. The table values characterize the Bondi-dominated limiting configurations and serve as reference points for the full coupled evolution, where the Eddington contribution can delay the approach to $\chi_{\rm fp}$.

\begin{figure}[htbp]
\centering
\includegraphics[width=0.84\linewidth]{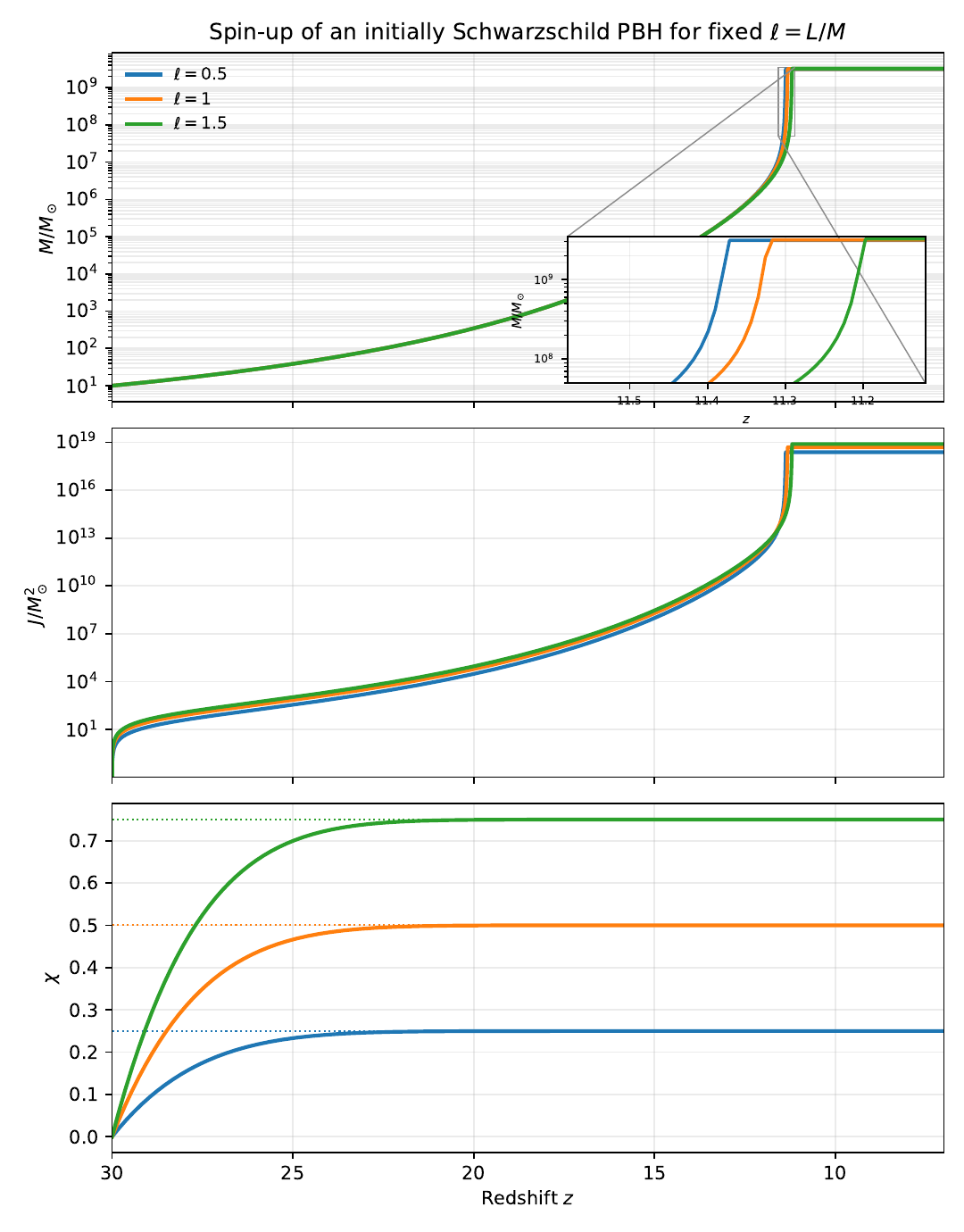}
\caption{Coupled mass–spin evolution of an initially Schwarzschild $10M_\odot$ PBH at $z_i=30$, for fixed inflow angular momenta $\ell=0.5,1,1.5$ and fiducial SIDM particle mass $m=10^{-2}\,\mathrm{eV}$. The evolution includes both Eddington growth and critical Kerr SIDM Bondi accretion, subject to the finite-halo supply prescription. The upper, middle, and lower panels show the black-hole mass, angular momentum, and dimensionless spin, respectively. The dotted lines indicate the Bondi-dominated limiting spins $\chi=\ell/2$. The inset resolves the rapid-growth interval and illustrates how the mass-growth history depends on the inflow angular momentum.
}
\label{fig:genericspinup}
\end{figure}

Figure~\ref{fig:genericspinup} illustrates the coupled mass–spin evolution for these representative fixed-$\ell$ benchmarks. The upper panel shows the black-hole mass, the middle panel the accumulated angular momentum, and the lower panel the dimensionless spin. The horizontal dotted lines indicate the Bondi-dominated limiting values $\chi=\ell/2$, rather than exact fixed points of the full evolution. The comparison isolates the feedback between inflow angular momentum, Kerr spin-up, and the critical SIDM accretion rate.

\clearpage

\subsection{The corotating analytic trajectory as a comparison}

The analytic $L=J/M$ line behaves very differently under secular evolution.  In the same Bondi-dominated benchmark $f_{\rm Bondi}=1$, one has $\ell=\chi$ and therefore
\begin{equation}
 \frac{\dd J}{\dd M}=L=\frac{J}{M},
\end{equation}
and therefore
\begin{equation}
 J(M)=J_i\frac{M}{M_i},\qquad
 \frac{J(M)}{M}=\frac{J_i}{M_i}={\rm const},\qquad
 \chi(M)=\chi_i\frac{M_i}{M}.
 \label{eq:special-spin-evolution}
\end{equation}
A Schwarzschild seed remains Schwarzschild, whereas an initially rotating seed becomes nearly nonrotating in the dimensionless sense after many orders of magnitude of mass growth.  Thus the $L=J/M$ branch is a valuable analytic benchmark but cannot generate a population of spinning SMBHs from initially nonspinning PBHs.

\begin{figure}[htbp]
\centering
\includegraphics[width=0.84\linewidth]{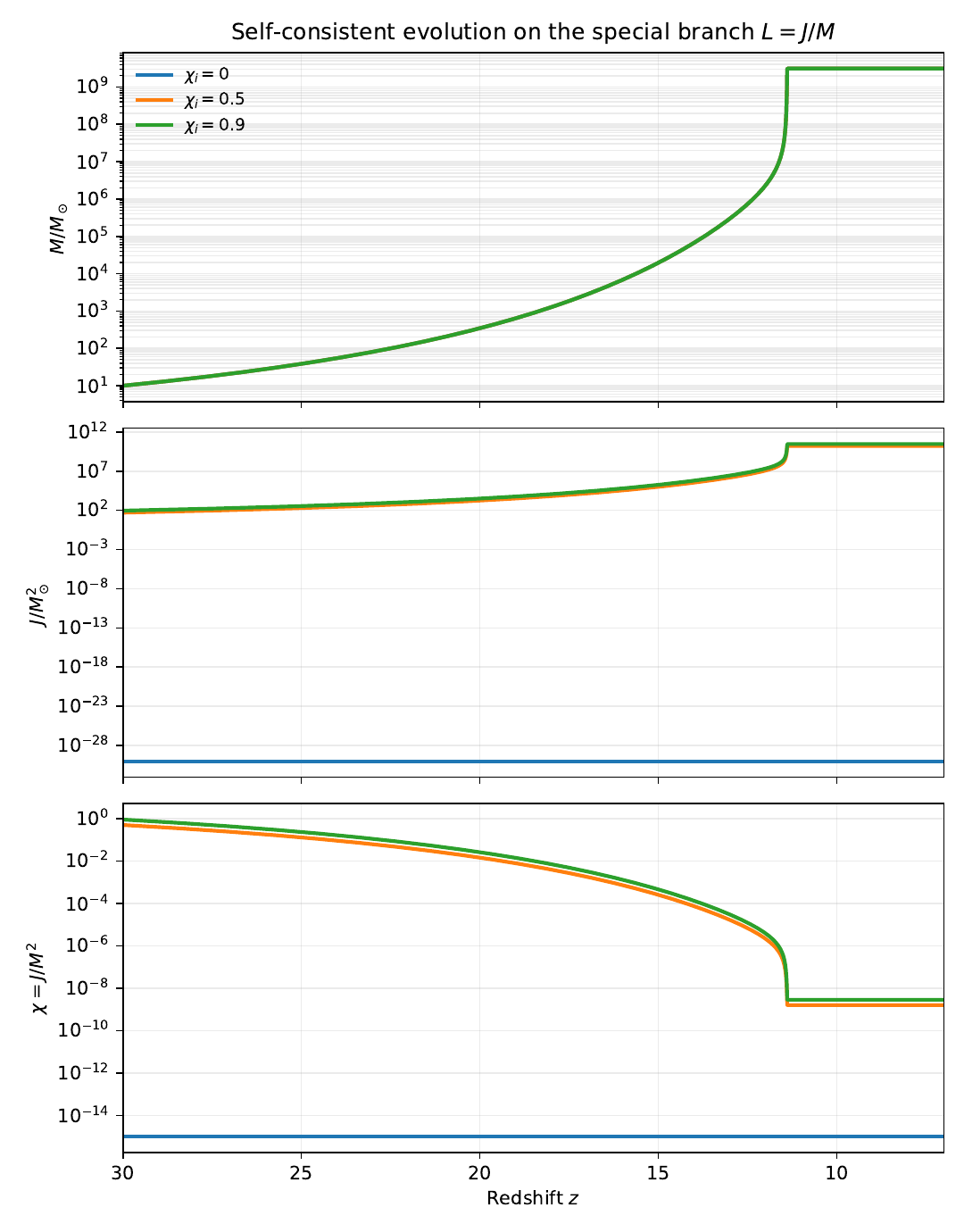}
\caption{Self-consistent evolution on the special $L=J/M$ branch for three initial spins.  The ratio $J/M$ stays constant, so $J\propto M$ and $\chi=J/M^2\propto M^{-1}$.  The $\chi_i=0$ curve remains exactly nonrotating.  The spin dependence of the mass history is too small to be resolved on the logarithmic upper panel: relative to the $\chi_i=0$ trajectory, $|\Delta M|/M$ is about $1.3\times10^{-7}$ ($4.3\times10^{-7}$) for $\chi_i=0.5$ ($0.9$) at $z\simeq12$, and grows to only $8.3\times10^{-6}$ ($2.8\times10^{-5}$) just before the common supply cap at $z\simeq11.39$.  Once the same finite SIDM supply cap is reached, all three mass trajectories coincide.}
\label{fig:specialevol}
\end{figure}

Figure~\ref{fig:specialevol} illustrates this behavior using the benchmark of Sec.~\ref{sec:numerical-benchmark}.  The mass curves are almost indistinguishable because $\chi$ rapidly decreases and $\Fspin(\chi)\to1$.

\clearpage

A fixed-$\chi$ stationary comparison along the line $\ell=\chi$ should therefore not be confused with a secular trajectory generated by the same flux: the self-consistent trajectory moves toward smaller $\chi$ as the mass grows.

\subsection{From PBH seed profiles to high-redshift SMBH populations}

The single-seed trajectories above establish that the critical SIDM Bondi channel can drive a stellar-mass black hole into the supermassive regime.  The more direct observational question is population-based: JWST is probing a growing sample of massive and overmassive black holes at $z\simeq7$--$10$, so we ask how a population of stellar-mass PBH seeds is mapped by the same mechanism into a high-redshift mass--spin distribution.  The calculation below is therefore an illustrative population template for comparison with the JWST SMBH census, not a fit to the current data.

Once an angular-momentum history is specified, number conservation maps an initial PBH population into a joint final distribution in mass and dimensionless Kerr spin,
\begin{equation}
 p_f(M,\chi)=\int\dd M_i\,\dd\ell\,p_i(M_i,\ell)
 \delta[M-M_f(M_i,\ell)]\,
 \delta[\chi-\chi_f(M_i,\ell)].
 \label{eq:jointmap}
\end{equation}
This choice is more directly tied to the Kerr state than using $J_f$ itself, because $J_f=\chi_f M_f^2$ carries a strong trivial mass dependence.  The corresponding marginal distributions are $P_f(M)=\int\dd\chi\,p_f(M,\chi)$ and $P_f(\chi)=\int\dd M\,p_f(M,\chi)$.
The stationary Kerr calculation does not determine the astrophysical prior for $\ell$, so we show conditional distributions for fixed $\ell=0.5,1,1.5$.  For the seed mass spectrum we use the same three representative stellar-mass PBH profiles as in our previous work: a log-normal profile with $M_c=5M_\odot$ and $\sigma=0.5$, a broad power law $p_i\propto M_i^{-2}$ over $0.1M_\odot\leq M_i\leq100M_\odot$, and a critical-collapse-like gamma-law profile with $\gamma=0.36$ and characteristic mass $M_c=5M_\odot$.  These narrow, broad, and asymmetric profiles are benchmark seed populations, not uniquely selected PBH mass functions.  Each is evolved from $z_i=30$ to $z=7$ with Eq.~\eqref{eq:growth-generic} and the finite-supply prescription of Sec.~\ref{sec:numerical-benchmark}.

\begin{figure}[p]
\centering
\includegraphics[width=0.96\linewidth]{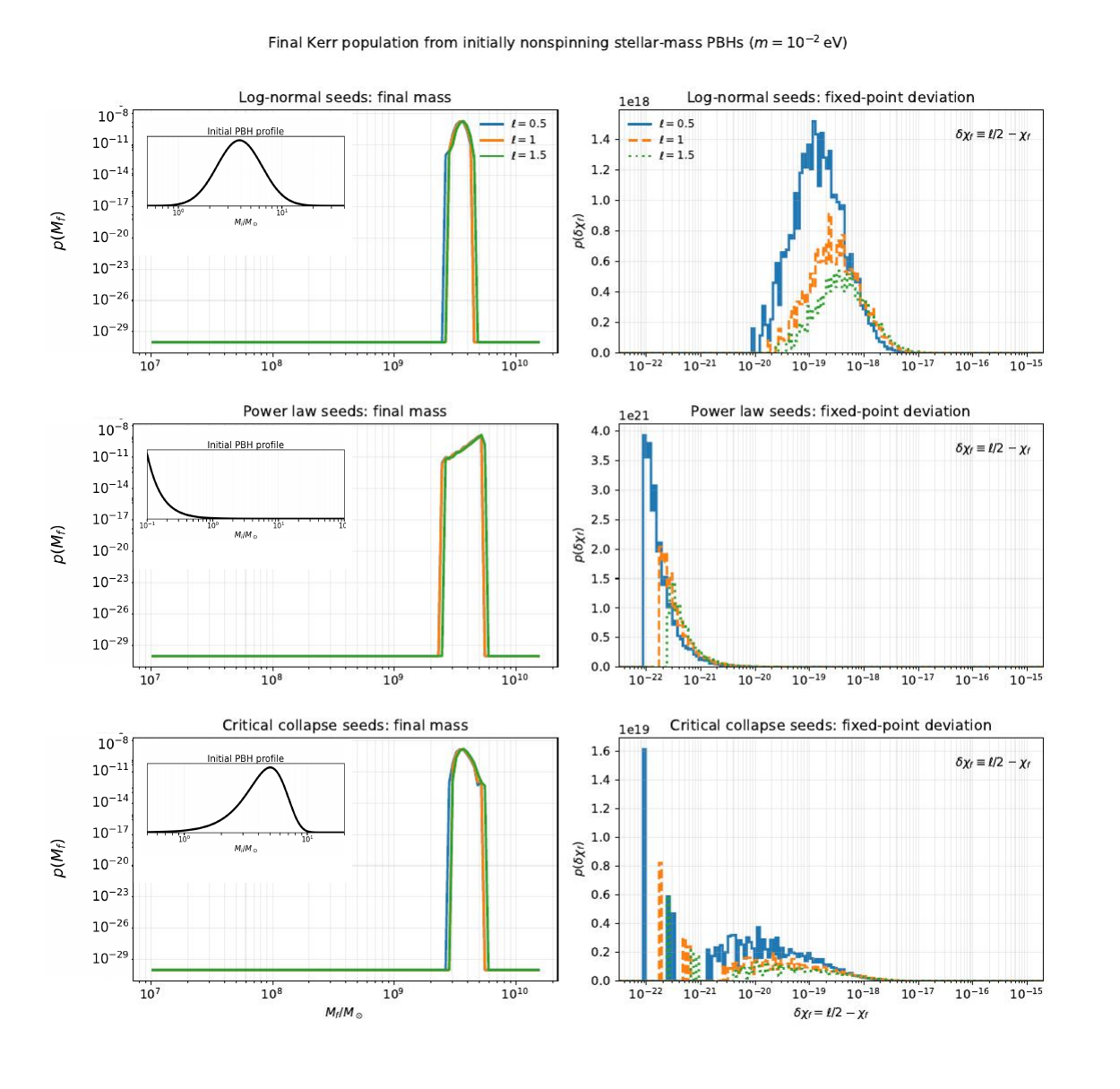}
\caption{Illustrative black-hole populations at $z=7$ from three initially nonspinning stellar-mass PBH seed profiles evolved through the critical Kerr SIDM Bondi channel for $m=10^{-2}\,\mathrm{eV}$ and fixed $\ell=0.5,1,1.5$. Left panels: final mass distributions; the insets show the corresponding normalized initial PBH seed profiles, common to all three $\ell$ values in each row. Right panels: probability densities for the fixed-point deviation $\delta\chi_f\equiv \ell/2-\chi_f$ on a logarithmic horizontal axis. The rows correspond to log-normal, power-law, and critical-collapse-like seed profiles.}
\label{fig:populations}
\end{figure}
\clearpage

Two features are robust within this illustrative model.  First, because the final mass is ultimately supply-limited, the mass pile-up remains in the supermassive range and depends only moderately on $\ell$.  Thus, rapid $M^2$ growth and the finite-halo cap can compress substantially different stellar-mass seed profiles into high-redshift SMBH populations with similar characteristic masses, providing the population-level version of the formation mechanism developed in this paper.  Second, for initially nonspinning seeds, the fixed-$\ell$ relation $\chi_f=(\ell/2)[1-(M_i/M_f)^2]$ drives the supermassive descendants extremely close to the spin fixed point $\chi_f=\ell/2$.  To resolve the small residual memory of the seed population, the right column of Fig.~\ref{fig:populations} therefore plots the positive deviation $\delta\chi_f\equiv\ell/2-\chi_f=(\ell/2)(M_i/M_f)^2$ on a logarithmic horizontal axis.  This shows that $\ell$ essentially fixes the absolute spin, while the minute departure from the fixed point still carries information about the seed distribution: broad power-law seeds produce the widest tail in $\delta\chi_f$, whereas log-normal and critical-collapse-like seeds remain more concentrated.  The final mass and spin distributions thus carry complementary information: the mass function primarily probes the growth efficiency and halo supply, whereas the dimensionless-spin outcome is controlled mainly by the angular momentum delivered by the SIDM flow, with only a tiny residual imprint of the seed masses.  A quantitative comparison with the JWST population would also require the seed abundance, halo occupation, selection function, and observational mass and spin uncertainties, which are beyond the scope of this illustrative calculation.

\section{Discussion and outlook}\label{sec:discussion}

The main result is that the critical SIDM growth mechanism proposed in our previous work \cite{ChenLin2026} survives the simultaneous inclusion of Kerr spin and fluid angular momentum.  The $\phi^4$ equation of state and the regular approach to a critical sonic point control the cancellation of the asymptotic halo density; therefore, it persists throughout a broad two-parameter domain rather than only along one analytically convenient trajectory.  Rotation changes the sonic radius and the dimensionless coefficient multiplying the $M^2$ growth law, but it does not generically destroy the rapid-growth channel.

A second result is that the most direct microphysics--astrophysics matching from the Schwarzschild problem is completely insensitive to the angular momenta.  Fixing the quartic coupling by the midpoint of the SIDM self-scattering window gives $\rhob\propto m^{5/2}$, while the finite-halo boundary condition gives $\zeta\simeq\rhodm/(8\rhob)$.  Hence $\zeta\propto\rhodm m^{-5/2}$, with no dependence on either $\chi$ or $\ell$.  This clean separation is useful conceptually: the outer halo and particle physics determine how close the flow is to the asymptotic critical limit, whereas Kerr dynamics determines the sonic geometry and multiplies the universal $M^2$ rate by $\Psic(\chi,\ell)$.  The coefficient $\Bcrit_{\chi\ell}$ only characterizes the local slope with which the sonic solution approaches its critical root as $\zeta\to0$; it does not modify the fundamental $\zeta$--$m$ relation.

The generic parameter space is naturally organized by the pair $(\chi,\ell)$.  The physical outer critical branch terminates precisely at the prograde and retrograde marginally bound Kerr angular momenta.  This gives a simple physical interpretation of the domain boundary: beyond the corresponding capture threshold, a quasi-spherical inviscid flow cannot pass smoothly through the centrifugal barrier without additional transport or multidimensional dynamics.  Within the domain, the exact $\ell=0$ solution shows that frame dragging by itself need not suppress the critical rate; the rate remains exactly equal to its Schwarzschild value even though the sonic location shifts.  The rate modification is therefore a joint effect of geometry and inflow angular momentum.

This distinction is important for the SMBH-assembly problem that motivates the paper.  The same SIDM inflow that changes the mass-growth coefficient also transports angular momentum.  Generic co-rotating flow can spin up an initially Schwarzschild stellar-mass seed as it grows, and the fixed-$\ell$ examples show how different angular-momentum histories yield distinct final Kerr configurations.  In the simple supply-limited model used here, the final mass distribution depends only moderately on $\ell$ because the host halo ultimately caps the available SIDM, whereas the dimensionless-spin distribution separates much more strongly between different inflow angular momenta.  If rapid SIDM growth contributes to the high-redshift black-hole population uncovered by JWST, the spin demographics would therefore provide information complementary to the mass function.  Turning this qualitative statement into an observational prediction will require a realistic distribution of SIDM angular momentum and, eventually, measurements or indirect constraints on the spins of early SMBHs.

The revised interpretation of the Bondi outer boundary also sharpens the role of the ambient density.  We now normalize $\rho_\infty$ to the cosmological SIDM density times an explicit environmental overdensity $\delta_{\rm env}$, rather than to the substantially larger fiducial value used in our previous work.  This lowers the density-derived criticality bound on the scalar mass by roughly two orders of magnitude in $m$ (and by several orders in density), but it does not lower the mass scale required for rapid cosmological growth, because the critical coefficient remains $K_0\propto m^{5/2}$ and is independent of $\rho_\infty$.  Thus $m\sim10^{-2}\,\mathrm{eV}$ remains a useful growth benchmark and is, with the revised density, extremely deep in the critical regime.  The main quantitative change at fixed $m$ is instead the larger Bondi radius, which makes the treatment of the finite halo environment relevant earlier and strengthens the need for a more realistic outer-halo matching in future work.

The matched corotating $L=J/M$ family should be viewed in this broader context.  It remains useful because many quantities are analytic and it supplies stringent checks on the generic formulas: the critical sonic radius moves inward with spin, the rate is suppressed by at most about $21\%$, and the fixed-growth scalar-mass comparison shifts only at the ten-percent level, while $\Bcrit_\chi$ provides an analytic diagnostic of the local sonic response.  Neither quantity changes the universal outer-boundary $\zeta$--$m$ relation.  Dynamically, however, this line is special.  It keeps the ratio $J/M$ constant, so the dimensionless spin $\chi=J/M^2$ decreases as the hole grows.  It therefore cannot generate a spinning SMBH population from initially nonrotating seeds, which is why the generic $(\chi,\ell)$ analysis is essential for the formation problem.

Several limitations should remain explicit.  The present flow is stationary, inviscid, adiabatic, equatorial, and quasi-spherical, with an effective $4\pi$ normalization.  We neglect viscosity, magnetic fields, the self-gravity of the inner flow, nonaxisymmetric structure, and shocks or outflows.  The broader Kerr and low-angular-momentum literature shows that centrifugal barriers can produce multiple critical points, thick structures, shocks, and strongly nonspherical accretion channels \cite{Goswami2007,Das2012,ShaikhDas2018,DasNandi2019,Sukova2017,Mach2018,Aguayo2021,Dihingia2025}.  Our outer critical branch is therefore best understood as a clean analytic sector whose global realization should be tested with two- and three-dimensional GR hydrodynamics.  The population calculations are likewise conditional examples rather than a prediction for the prior distribution of $\ell$.

Taken together, these results show that the critical SIDM Bondi mechanism identified in the Schwarzschild problem remains viable for rotating black holes with generic inflow angular momentum.  Kerr rotation changes the sonic geometry, the critical-rate coefficient, and the spin evolution, but preserves the characteristic $M^2$ growth that can rapidly build supermassive black holes from stellar-mass seeds.  The resulting mass--spin dependence therefore provides a direct Kerr extension of our previous SIDM growth scenario and retains its connection between dark-matter microphysics and the early massive black-hole population revealed by JWST.

\begin{acknowledgments}
We thank Van Que Tran for helpful discussions, especially for pointing out that the ambient dark matter density in the previous paper is a few orders of magnitude larger than the conventional value. We acknowledge the use of GPT-5.6 Sol in deriving some results and preparing the manuscript.  The work of F.L.L. is supported by Taiwan's National Science and Technology Council (NSTC) through Grant Nos.~115-2112-M-003-009-MY3 and 112-2112-M-003-006-MY3. 
\end{acknowledgments}

\appendix

\section{Detailed derivation of stationary Bondi flow on Kerr}\label{app:kerrflow}

This appendix gives the full stationary-flow derivation using the dimensionless variables introduced at the beginning of the main text.  Thus
\[
 \tau=t/M,\qquad x=r/M,\qquad \chi=J/M^2,\qquad \ell=L/M,
\]
and the dimensionless Kerr functions are $\Sigma=x^2+\chi^2\cos^2\theta$ and $\Delta=x^2-2x+\chi^2$.  We also keep the main-text inward convention $u_{\rm in}\equiv-u^x>0$, where $u^x=\dd x/\dd\hat\lambda$ is the signed coordinate component of the four-velocity.  We use the same radial-subscript convention as in the main text: $c_{\rm s}$ is the local sound speed; a separate subscript $s$ means evaluation at the sonic point $x_s$; a subscript $c$ means evaluation on the critical sonic solution $x_c$; and $\infty$ denotes the asymptotic outer-halo state.  We denote the asymptotic squared sound speed by $\zeta$, as defined in Eq.~\eqref{eq:zeta-def}.  The symbol $\sigma$ below denotes specific entropy and is unrelated to the sonic subscript.

\subsection{Dimensionless Kerr geometry and Killing currents}

For the metric~\eqref{eq:kerr-dimless}, the nonzero inverse components are
\begin{align}
 g^{\tau\tau}&=-\frac{(x^2+\chi^2)^2-\chi^2\Delta\sin^2\theta}{\Delta\Sigma},\\
 g^{\tau\phi}&=-\frac{2\chi x}{\Delta\Sigma},\\
 g^{\phi\phi}&=\frac{\Delta-\chi^2\sin^2\theta}{\Delta\Sigma\sin^2\theta},\\
 g^{xx}&=\frac{\Delta}{\Sigma},\qquad g^{\theta\theta}=\frac1\Sigma.
\end{align}
The stationary and axial Killing fields are $\xi_{(\tau)}=\partial_\tau$ and $\xi_{(\phi)}=\partial_\phi$.  Stress-tensor conservation gives the energy current and the dimensional angular-momentum current
\begin{equation}
 j_E^a=-T^a{}_{\tau},\qquad
 j_J^a=M T^a{}_{\phi}.
\end{equation}

\subsection{Euler equation and streamline integrals}

Rest-mass conservation and stress-tensor conservation are
\begin{equation}
 \nabla_a(\rho_0u^a)=0,
 \qquad
 \nabla_a T^{ab}=0.
\end{equation}
Contracting the latter with $u_b$ and combining with the former yields
\begin{equation}
 u^a\nabla_a\rho=h\,u^a\nabla_a\rho_0,
\end{equation}
so along an adiabatic streamline
\begin{equation}
 \left(\frac{\partial\rho}{\partial\rho_0}\right)_\sigma=h.
\end{equation}
The projection orthogonal to $u^a$ gives, for isentropic flow,
\begin{equation}
 u^a\nabla_a(hu_b)+\nabla_b h=0.
 \label{eq:euler-app}
\end{equation}
If $\xi^b$ is a Killing field respected by the fluid, then $u^a\nabla_a(hu_b\xi^b)=0$.  Applying this to $\partial_\tau$ and $\partial_\phi$ gives
\begin{equation}
 E=-hu_\tau,\qquad h u_\phi=E\ell,
\end{equation}
which is equivalent to Eq.~\eqref{eq:ELmain}, with the dimensional angular momentum $L=M\ell$.

Writing $u^\phi=\Omega u^\tau$, where $\Omega=d\phi/d\tau$ is dimensionless, the covariant components are
\begin{equation}
 u_\tau=u^\tau(g_{\tau\tau}+g_{\tau\phi}\Omega),\qquad
 u_\phi=u^\tau(g_{\tau\phi}+g_{\phi\phi}\Omega).
\end{equation}
Solving $\ell=-u_\phi/u_\tau$ gives Eq.~\eqref{eq:omega-main}.  On the equatorial plane,
\begin{equation}
 \Omega(x)=\frac{2\chi+(x-2)\ell}
 {\chi^2(x+2)-2\chi\ell+x^3}.
 \label{eq:Omega-explicit}
\end{equation}

\subsection{Normalization and Bernoulli equation}

Define the poloidal norm in the dimensionless geometry by
\begin{equation}
 u_{\rm pol}^2=g_{xx}u_{\rm in}^2+g_{\theta\theta}(u^\theta)^2,
\end{equation}
where $u_{\rm in}=-u^x>0$ is the inward radial speed.
The normalization becomes
\begin{equation}
 g_{\tau\tau}^{\rm eff}(u^\tau)^2+u_{\rm pol}^2=-1,
 \qquad
 g_{\tau\tau}^{\rm eff}=g_{\tau\tau}+2g_{\tau\phi}\Omega+g_{\phi\phi}\Omega^2.
 \label{eq:norm-app}
\end{equation}
Using $E=-hu_\tau$ gives
\begin{equation}
 E^2=-h^2(1+u_{\rm pol}^2)\Gcal,
 \qquad
 \Gcal=\frac{(g_{\tau\tau}+g_{\tau\phi}\Omega)^2}
 {g_{\tau\tau}^{\rm eff}}.
 \label{eq:bern-app}
\end{equation}
On the equatorial plane, substitution of Eq.~\eqref{eq:Omega-explicit} gives the compact dimensionless form
\begin{equation}
 \Gcal(x)=-\frac{x\Delta(x)}{A(x;\chi,\ell)},
 \label{eq:Gexplicit}
\end{equation}
where $A$ is defined in Eq.~\eqref{eq:Adef}.  This form makes the role of the single Kerr function $\Delta(x)$ manifest.

At the asymptotic outer boundary one has $\Gcal_\infty\to-1$, and we impose $u_{{\rm pol},\infty}\to0$.  Evaluating the Bernoulli constant at the sonic point and at infinity therefore gives
\begin{equation}
 h_s^2(1+u_{{\rm pol},s}^2)(-\Gcal_s)=h_\infty^2.
 \label{eq:bern-endpoints}
\end{equation}

\subsection{Continuity and wind equation}

On the equatorial plane, $\sqrt{-g}=x^2$ for the dimensionless metric.  Since the signed coordinate component is $u^x=-u_{\rm in}$, stationarity and axisymmetry reduce the continuity equation to
\begin{equation}
 \partial_x(x^2\rho_0u_{\rm in})=0,
\end{equation}
which gives Eq.~\eqref{eq:restmassflux} after restoring the overall area factor $M^2$.  Along the future-directed dimensionless streamline, $\dd x/\dd\hat\lambda=-u_{\rm in}$, and
\begin{equation}
 \frac{\dd\rho_0}{\dd\hat\lambda}=-\rho_0\nabla_a u^a,
 \qquad
 \nabla_a u^a=\frac1{u_{\rm in}}\frac{\dd u_{\rm in}}{\dd\hat\lambda}-\frac{2u_{\rm in}}{x}.
\end{equation}
For adiabatic flow,
\begin{equation}
 \frac{\dd h}{\dd\hat\lambda}
 =-hc_{\rm s}^2\left(\frac1{u_{\rm in}}\frac{\dd u_{\rm in}}{\dd\hat\lambda}-\frac{2u_{\rm in}}{x}\right).
 \label{eq:dh-app}
\end{equation}
Since $u_{\rm pol}^2=g_{xx}u_{\rm in}^2$ in the equatorial reduction,
\begin{equation}
 \frac1{u_{\rm in}}\frac{\dd u_{\rm in}}{\dd\hat\lambda}
 =\frac1{u_{\rm pol}}\frac{\dd u_{\rm pol}}{\dd\hat\lambda}
 -\frac12\frac{\dd\ln g_{xx}}{\dd\hat\lambda}.
\end{equation}
Differentiating Eq.~\eqref{eq:bern-app} and using Eq.~\eqref{eq:dh-app} gives
\begin{equation}
 \frac{\dd u_{\rm pol}}{\dd\hat\lambda}=\frac{\Ncal}{\Dcal},
\end{equation}
with
\begin{align}
 \Dcal&=1+\frac{u_{\rm pol}^2c_{\rm s}^2}{g_{\tau\tau}^{\rm eff}(u^\tau)^2},\label{eq:Dapp}\\
 \Ncal&=\frac{u_{\rm pol}}{2}
 \left[\frac1{c_{\rm s}^2}\frac{\dd\ln\Gcal}{\dd\hat\lambda}
 +\frac{4u_{\rm in}}{x}+\frac{\dd\ln g_{xx}}{\dd\hat\lambda}\right].
 \label{eq:Napp}
\end{align}
The sonic point requires $\Dcal_s=\Ncal_s=0$.  From $\Dcal_s=0$ and Eq.~\eqref{eq:norm-app},
\begin{equation}
 u_{{\rm pol},s}^2=\frac{c_{{\rm s},s}^2}{1-c_{{\rm s},s}^2}.
\end{equation}
On the equatorial plane $g_{xx}=x^2/\Delta$, so
\begin{equation}
 u_{{\rm in},s}^2=\frac{c_{{\rm s},s}^2\Delta_s}
 {x_s^2(1-c_{{\rm s},s}^2)},
\end{equation}
which is Eq.~\eqref{eq:sonicDmain}.

\subsection{General sonic regularity condition}

Evaluating $\Ncal_s=0$ using $\Gcal=-x\Delta/A$ and $g_{xx}=x^2/\Delta$ gives directly
\begin{equation}
 c_{{\rm s},s}^2=\frac{n_s}{A_sB_s},
 \label{eq:css-dimensionful}
\end{equation}
where $A_s=A(x_s)$, $B_s=B(x_s)$, and $n_s=n(x_s)$ are the dimensionless functions defined in Eqs.~\eqref{eq:Adef}--\eqref{eq:ndef}.  Combining this with the denominator condition gives
\begin{equation}
 u_{{\rm in},s}^2=
 \frac{n_s}{x_s^2\left[(\chi-\ell)^2+x_s(\chi^2-\ell^2)+2x_s^3\right]}.
 \label{eq:ur-dimensionful}
\end{equation}
Thus the full hydrodynamic sonic conditions are dimensionless from the start; no intermediate dimensional $r_s$, $L$, or Kerr $\Delta(r)$ is required.

\subsection{Thermodynamic identities, Mach number, and the Bernoulli map}\label{app:thermo-bern}

It is useful to derive the thermodynamic relations that enter Eq.~\eqref{eq:cinf-generic-main}.  Write
\begin{equation}
 y\equiv c_{\rm s}^2.
\end{equation}
Equation~\eqref{eq:rhop-main} becomes
\begin{equation}
 \rho(y)=4\rhob\frac{(2-3y)y}{(1-3y)^2},
 \qquad
 p(y)=4\rhob\frac{y^2}{(1-3y)^2}.
 \label{eq:rho-p-y-app}
\end{equation}
For an isentropic perfect fluid, $\dd\rho=h\,\dd\rho_0$ and $h=(\rho+p)/\rho_0$, hence
\begin{equation}
 \frac{\dd\ln\rho_0}{\dd y}
 =\frac{\dd\rho/\dd y}{\rho+p}
 =\frac1y+\frac{1}{2(y-1)}-\frac{9}{2(3y-1)}.
 \label{eq:dlnrho0-y}
\end{equation}
Integration gives
\begin{equation}
 \rho_0=C\,
 \frac{y\sqrt{1-y}}{(1-3y)^{3/2}}.
\end{equation}
The cold limit has $\rho\simeq8\rhob y$ and must satisfy $h\to1$, so $C=8\rhob$.  Therefore
\begin{equation}
 \rho_0(y)=8\rhob\frac{y\sqrt{1-y}}{(1-3y)^{3/2}},
 \qquad
 h(y)=\sqrt{\frac{1-y}{1-3y}}.
 \label{eq:rho0-h-y}
\end{equation}
Taking the ratio of Eq.~\eqref{eq:rho0-h-y} between the sonic point and infinity immediately reproduces Eq.~\eqref{eq:rho0ratio-main}.

For completeness, the actual relativistic poloidal Mach number is defined from the local poloidal three-speed,
\begin{equation}
 v_{\rm pol}^2\equiv\frac{u_{\rm pol}^2}{1+u_{\rm pol}^2},
 \qquad
 {\rm Ma}^2\equiv\frac{v_{\rm pol}^2}{c_{\rm s}^2}
 =\frac{u_{\rm pol}^2}{c_{\rm s}^2(1+u_{\rm pol}^2)}.
 \label{eq:mach-app}
\end{equation}
Define $y_s\equiv y(x_s)=c_{{\rm s},s}^2$.  The sonic denominator condition gives
$u_{{\rm pol},s}^2=y_s/(1-y_s)$ and hence ${\rm Ma}_s=1$.  This $\mathrm{Ma}$ should not be confused with the auxiliary sonic-point quantity $\Mcal_s$ appearing in Eq.~\eqref{eq:cinf-generic-main}.

We now derive that equation.  We define the dimensionless radial coordinate $x$ in Eq.~\eqref{eq:dimensionless}, and $x_s$ denotes its sonic-point value.  We similarly define $A_s=A(x_s)$ and $\Delta_s=\Delta(x_s)$.  On the equatorial plane Eq.~\eqref{eq:Gexplicit}, evaluated at $x=x_s$, gives
\begin{equation}
 -\Gcal_s=\frac{x_s\Delta_s}{A_s}.
 \label{eq:Gsonic-app}
\end{equation}
We define the asymptotic sound-speed variable $\zeta$ in Eq.~\eqref{eq:zeta-def}.  Here $h_s\equiv h(x_s)$ and $h_\infty\equiv h(x\to\infty)$.  From Eq.~\eqref{eq:rho0-h-y},
\begin{equation}
 \frac{h_s^2}{h_\infty^2}
 =\frac{1-y_s}{1-3y_s}\,
  \frac{1-3\zeta}{1-\zeta}.
\end{equation}
Using $1+u_{{\rm pol},s}^2=1/(1-y_s)$ in the endpoint Bernoulli relation~\eqref{eq:bern-endpoints} then gives
\begin{equation}
 \frac{x_s\Delta_s}{A_s}
 =\frac{(1-3y_s)(1-\zeta)}{1-3\zeta}.
 \label{eq:bern-cs-app}
\end{equation}
Rearranging,
\begin{equation}
 \frac{A_s}{x_s\Delta_s}(1-3y_s)
 =\frac{1-3\zeta}{1-\zeta}.
\end{equation}
The left-hand side is precisely the auxiliary sonic-point quantity
\begin{equation}
 \Mcal_s\equiv\frac{A_s}{x_s\Delta_s}(1-3y_s).
\end{equation}
Thus $\Mcal_s(1-\zeta)=1-3\zeta$, and solving for $\zeta$ yields
\begin{equation}
 \zeta=\frac{\Mcal_s-1}{\Mcal_s-3},
\end{equation}
which is Eq.~\eqref{eq:cinf-generic-main}.

Finally, Eq.~\eqref{eq:rhoinf-smallz} follows directly from Eq.~\eqref{eq:rho-p-y-app}.  Expanding at $y=\zeta\ll1$ gives
\begin{align}
 \rho_\infty&=8\rhob\zeta+36\rhob\zeta^2+O(\zeta^3),\\
 p_\infty&=4\rhob\zeta^2+24\rhob\zeta^3+O(\zeta^4),\\
 \rho_{0,\infty}&=8\rhob\zeta\left[1+4\zeta+O(\zeta^2)\right].
 \label{eq:asymp-expanded-app}
\end{align}
Hence $p_\infty/\rho_\infty=O(\zeta)$ and $\rho_{0,\infty}/\rho_\infty=1+O(\zeta)$, justifying the pressureless asymptotic identification used in the main text.

\subsection{Rest-mass, energy, and angular-momentum rates}

Using Eq.~\eqref{eq:rho0ratio-main} at the sonic point together with Eq.~\eqref{eq:sonicDmain}, one obtains
\begin{align}
 \dot M_{\rm rest}={}&4\pi M^2\rho_{0,\infty}
 \frac{(1-\zeta)(c_{{\rm s},s}^2)^{3/2}}
 {\zeta}
 \frac{A_s^{3/2}}{x_s^{1/2}\Delta_s},
 \label{eq:mdot-app}
\end{align}
where $A_s$, $x_s$, and $\Delta_s$ are sonic-point quantities as defined above.  This is Eq.~\eqref{eq:mdot-generic}.  Recall that this is the conserved \emph{rest-mass} flux.  The corresponding Bondi-induced changes in the Kerr parameters follow from Eq.~\eqref{eq:energyJflux}: $\dot M_{\rm Bondi}=E\dot M_{\rm rest}$ and $\dot J=EL\dot M_{\rm rest}$.  Thus the stationary continuity equation determines how much rest mass crosses a surface, while the stress-energy flux determines how much black-hole mass-energy and angular momentum are gained through SIDM Bondi accretion.

\section{Details of the Kerr critical regime}\label{app:critical}

We keep the same dimensionless notation as in the main text and Appendix~\ref{app:kerrflow}: $x_s$ labels an ordinary sonic point, $Q_s$ means $Q$ evaluated there, $Q_c$ denotes the corresponding quantity on the critical sonic solution $x_s=x_c$, and $Q_\infty$ denotes the outer-halo boundary value.  In particular, $A_s=A(x_s)$, $B_s=B(x_s)$, $n_s=n(x_s)$, and $\Delta_s=\Delta(x_s)$.

\subsection{Criticality, the quartic, and the physical root}\label{app:quartic-domain}

Along the family of regular sonic points, the dimensionless sonic coordinate $x_s$ is the natural parameter.  We therefore regard
\begin{equation}
 y_s(x_s)\equiv c_{{\rm s},s}^2,
 \qquad
 \zeta(x_s)=\frac{\Mcal_s(x_s)-1}{\Mcal_s(x_s)-3}
\end{equation}
as functions of $x_s$, with
\begin{equation}
 y_s=\frac{n_s}{A_sB_s},\qquad
 \Mcal_s=\frac{A_s}{x_s\Delta_s}(1-3y_s).
\end{equation}
A critical point with a cold asymptotic state has $\zeta(x_c)=0$, where $x_c$ is the critical value of $x_s$.  We define $y_c\equiv y_s(x_c)=c_{{\rm s},c}^2$ and $\Mcal_c\equiv\Mcal_s(x_c)=1$.  Before setting $x_s=x_c$, the condition $\Mcal_s=1$ is
\begin{equation}
 A_s\left(1-\frac{3n_s}{A_sB_s}\right)=x_s\Delta_s.
\end{equation}
Multiplying by $B_s$ gives the particularly useful polynomial form
\begin{equation}
 B_s(A_s-x_s\Delta_s)-3n_s=0.
 \label{eq:quartic-intermediate}
\end{equation}
Substituting Eqs.~\eqref{eq:Adef}--\eqref{eq:ndef} and expanding Eq.~\eqref{eq:quartic-intermediate} gives
\begin{align}
 0={}&x_s^4+(\ell^2-6)x_s^3+(10\chi\ell-5\ell^2)x_s^2\nonumber\\
 &+\left[6(\chi-\ell)^2-\chi^2\ell^2\right]x_s
 -\chi^2(\chi-\ell)^2,
\end{align}
which is Eq.~\eqref{eq:criticalquartic}.  The physical critical root is denoted $x_s=x_c$.

The algebraic quartic generally has several real or complex roots.  The outer sonic point must satisfy the physical filters in Eq.~\eqref{eq:physical-root}.  The Kerr horizons obey $\Delta=0$, hence
\begin{equation}
 x_\pm=1\pm\sqrt{1-\chi^2},
 \qquad x_+=\frac{r_+}{M}.
 \label{eq:xpm-app}
\end{equation}
Thus $x_c>x_+$ is simply the requirement that the sonic transition occur outside the future event horizon.  The interval $0<y_c<1/3$ is the physical $\phi^4$ branch: positivity excludes an imaginary sound speed, while the upper endpoint is the stiff high-density limit at which the denominators in Eqs.~\eqref{eq:rhop-main} and \eqref{eq:rho0-h-y} become singular.  At the critical sonic point outside the horizon, $x_c\Delta_c>0$; because $-\Gcal_c=x_c\Delta_c/A_c$, positivity of the Bernoulli factor requires $A_c>0$.  The final condition $E(1-\Omega_H\ell)>0$ is the usual future-horizon energy condition in the dimensionless convention of Eq.~\eqref{eq:horizon-energy}.

\subsection{Derivation of the generic existence interval}\label{app:domain-derivation}

At an endpoint of the outer critical branch, the sonic sound speed tends to zero, $y_c\to0$.  From $y_s=n_s/(A_sB_s)$ this requires
\begin{equation}
 n_c\equiv n(x_c;\chi,\ell)=0.
 \label{eq:endpoint-n0}
\end{equation}
At the same time $\Mcal_c=1$ reduces to
\begin{equation}
 A_c=x_c\Delta_c,
\end{equation}
which, after inserting $A_c=A(x_c)$, is
\begin{equation}
 2x_c^2-\ell^2x_c+2(\chi-\ell)^2=0.
 \label{eq:endpoint-A}
\end{equation}
Eliminating $x_c$ between Eqs.~\eqref{eq:endpoint-n0} and \eqref{eq:endpoint-A} gives the resultant
\begin{align}
 0={}&(\chi-\ell)^2
 (4\chi-\ell^2-4\ell)
 (4\chi+\ell^2-4\ell)\nonumber\\
 &\times(\chi\ell^2+4\chi-4\ell)^2.
 \label{eq:endpoint-resultant}
\end{align}
The factors contain several algebraic degeneracies, but the physical \emph{outer} branch is fixed by $x_c>x_+$ and by continuity with Schwarzschild, where the critical capture interval is $-4<\ell<4$.  The relevant roots are therefore
\begin{equation}
 \ell_-=-2\left(1+\sqrt{1+\chi}\right),
 \qquad
 \ell_+=2\left(1+\sqrt{1-\chi}\right).
 \label{eq:ell-end-app}
\end{equation}
The alternative small-magnitude roots of the same quadratic factors give candidate radii $x_c\le x_+$ (indeed they lie deep inside the outer horizon for nonextremal Kerr), while the factor $\chi\ell^2+4\chi-4\ell$ produces a horizon or inner-root degeneracy.  The factor $\ell=\chi$ is the regular special family and is not a boundary.

The corresponding outer endpoint radii follow by substituting Eq.~\eqref{eq:ell-end-app} into Eq.~\eqref{eq:endpoint-A}.  With $\eta_+\equiv\sqrt{1-\chi}$ and $\eta_-\equiv\sqrt{1+\chi}$,
\begin{equation}
 x_{\rm end}^+=(1+\eta_+)^2
 =2-\chi+2\sqrt{1-\chi},
\end{equation}
\begin{equation}
 x_{\rm end}^-=(1+\eta_-)^2
 =2+\chi+2\sqrt{1+\chi}.
\end{equation}
These are exactly the prograde and retrograde marginally bound Kerr radii.  Between these two analytic endpoints, direct continuation of the largest root of Eq.~\eqref{eq:criticalquartic} preserves all filters in Eq.~\eqref{eq:physical-root}; outside them the continuation either yields $y_c\le0$ or fails the outer-horizon/capture conditions.  This establishes the physical interval quoted in Eq.~\eqref{eq:ell-domain}.  At extremal prograde spin, the endpoint $x_{\rm end}^+=x_+=1$ lies on the horizon, so the interval is open.

\subsection{Near-critical response for a simple root}\label{app:nearcritical}

For a simple critical root, define $\delta x_s\equiv x_s-x_c$.  A prime in this subsection means differentiation with respect to the sonic coordinate $x_s$.  Then
\begin{equation}
 \zeta(x_s)=\zeta'_c\,\delta x_s+O(\delta x_s^2).
\end{equation}
Differentiating $\zeta=(\Mcal_s-1)/(\Mcal_s-3)$ and then setting $\Mcal_c=1$ gives
\begin{equation}
 \zeta'_c=-\frac12\Mcal'_{s,c},
\end{equation}
where $\Mcal'_{s,c}\equiv(\dd\Mcal_s/\dd x_s)_{x_s=x_c}$.  Likewise,
\begin{equation}
 y_s(x_s)=y_c+y'_{s,c}\,\delta x_s+O(\delta x_s^2),
\end{equation}
with $y_s\equiv c_{{\rm s},s}^2$, $y_c\equiv c_{{\rm s},c}^2$, and $y'_{s,c}\equiv(\dd y_s/\dd x_s)_{x_s=x_c}$.  Eliminating $\delta x_s$ directly, without introducing any additional critical parameter, gives
\begin{equation}
 \zeta=\Bcrit_{\chi\ell}(y_s-y_c)
 +O\!\left[(y_s-y_c)^2\right],
 \qquad
 \Bcrit_{\chi\ell}=-\frac{\Mcal'_{s,c}}{2y'_{s,c}}.
 \label{eq:Bgen}
\end{equation}

To make the coefficient explicit, write
\begin{equation}
 y_s=\frac{n_s}{A_sB_s},\qquad
 \Mcal_s=\frac{A_sB_s-3n_s}{x_s\Delta_sB_s}.
\end{equation}
At the critical root $\Mcal_c=1$, so $A_cB_c-3n_c=x_c\Delta_cB_c$.  Differentiation therefore gives
\begin{equation}
 y'_{s,c}=\frac{n'_cA_cB_c-n_c(A'_cB_c+A_cB'_c)}{A_c^2B_c^2},
 \label{eq:yprime-explicit}
\end{equation}
\begin{equation}
 \Mcal'_{s,c}=
 \frac{A'_cB_c+A_cB'_c-3n'_c-
 [(\Delta_c+x_c\Delta'_c)B_c+x_c\Delta_cB'_c]}
 {x_c\Delta_cB_c}.
 \label{eq:Mprime-explicit}
\end{equation}
Combining Eqs.~\eqref{eq:yprime-explicit} and \eqref{eq:Mprime-explicit} yields
\begin{equation}
 \Bcrit_{\chi\ell}
 =\frac{A_c^2B_c}{2x_c\Delta_c}
 \frac{
 [ (\Delta_c+x_c\Delta'_c)B_c+x_c\Delta_c B'_c]
 -[A'_cB_c+A_cB'_c-3n'_c]
 }
 {n'_cA_cB_c-n_c(A'_cB_c+A_cB'_c)}.
 \label{eq:Bexplicit-app}
\end{equation}
The required derivatives contain no implicit differentiation,
\begin{equation}
\begin{aligned}
 A'_c&=\chi^2-\ell^2+3x_c^2,\qquad
 B'_c=4x_c-3,\qquad
 \Delta'_c=2(x_c-1),\\
 n'_c&=-4(\chi-\ell)^2
 +4x_c(\chi-2\ell)(\chi-\ell)
 -3\ell^2x_c^2+4x_c^3.
\end{aligned}
\label{eq:Bderivs-app}
\end{equation}
Equations~\eqref{eq:Bexplicit-app} and \eqref{eq:Bderivs-app} are the explicit forms quoted in the main text.  The coefficient $\Bcrit_{\chi\ell}$ is a local sonic-response quantity only; the direct outer-boundary relation between $\zeta$ and the scalar mass is already fixed by Sec.~\ref{sec:zeta-microphysics} and contains no $\Bcrit_{\chi\ell}$.

\subsection{Cancellation of the critical pole and derivation of the generic rate}\label{app:critical-pole}

Equation~\eqref{eq:mdot-app} may be written directly in terms of $\zeta$ as
\begin{equation}
 \dot M_{\rm rest}=4\pi M^2\rho_{0,\infty}
 \frac{1-\zeta}{\zeta}
 y_s^{3/2}\frac{A_s^{3/2}}{x_s^{1/2}\Delta_s}.
 \label{eq:mdot-zeta-app}
\end{equation}
From Eq.~\eqref{eq:asymp-expanded-app},
\begin{equation}
 \rho_{0,\infty}=8\rhob\zeta[1+4\zeta+O(\zeta^2)].
\end{equation}
Hence the apparently singular prefactor has the finite limit
\begin{equation}
 \lim_{\zeta\to0}
 \rho_{0,\infty}\frac{1-\zeta}{\zeta}
 =8\rhob.
 \label{eq:pole-limit-app}
\end{equation}
At a regular critical root, $y_s\to y_c=c_{{\rm s},c}^2$, $A_s\to A_c$, $x_s\to x_c$, and $\Delta_s\to\Delta_c$.  Applying Eq.~\eqref{eq:pole-limit-app} to Eq.~\eqref{eq:mdot-zeta-app} therefore gives
\begin{equation}
 \dot M_{{\rm rest},{\rm crit}}
 =32\pi\rhob M^2
 \frac{(c_{{\rm s},c}^2)^{3/2}A_c^{3/2}}
 {x_c^{1/2}\Delta_c}.
 \label{eq:generic-rate-raw-app}
\end{equation}
It is convenient to normalize the dimensionless geometric factor so that the Schwarzschild result is $\Psic=1$:
\begin{equation}
 \Psic(\chi,\ell)\equiv\frac12
 \frac{(c_{{\rm s},c}^2)^{3/2}A_c^{3/2}}
 {x_c^{1/2}\Delta_c}.
 \label{eq:Psi-derived-app}
\end{equation}
Then Eq.~\eqref{eq:generic-rate-raw-app} becomes
\begin{equation}
 \dot M_{{\rm rest},{\rm crit}}=64\pi\Psic(\chi,\ell)\rhob M^2,
\end{equation}
which proves Eqs.~\eqref{eq:generic-rate} and \eqref{eq:Psi-generic}.  Since $E=h_\infty+O(u_{{\rm pol},\infty}^2)=1+O(\zeta)$, Eq.~\eqref{eq:energyJflux} implies $\dot M_{\rm Bondi}=\dot M_{\rm rest}[1+O(\zeta)]$: the Bondi-induced black-hole mass-energy rate agrees with the conserved rest-mass rate only at leading critical order.  The cancellation also makes clear why $\Bcrit_{\chi\ell}$ does not enter the universal rate: $\Bcrit_{\chi\ell}$ determines how the halo boundary condition approaches $\zeta=0$, whereas the factor of $\zeta$ itself cancels locally in Eq.~\eqref{eq:pole-limit-app}.

\subsection{Special branch $L=J/M$ ($\ell=\chi$): exact formulas and near-critical expansion}\label{app:special-expansion}

For $\ell=\chi$, the generic sonic expression becomes Eq.~\eqref{eq:css-special} and the Bernoulli map can be written in the rational form
\begin{equation}
 \zeta=\frac{P_\chi(x_s)}{Q_\chi(x_s)},
 \label{eq:cinf-special-app}
\end{equation}
where $P_\chi$ is defined in Eq.~\eqref{eq:Pchi} and
\begin{equation}
 Q_\chi(X)=3\chi^4+5\chi^2X(X-3)+X^2(4X^2-15X+18).
 \label{eq:Qchi-app}
\end{equation}
Substituting the same specialization into the general flux formula gives the exact precritical rest-mass rate
\begin{equation}
 \dot M_{\rm rest}=
 8\pi M^2\rho_{0,\infty}
 \frac{x_s^4(x_s-\chi^2)^{3/2}}
 {P_\chi(x_s)\sqrt{\chi^2+x_s(2x_s-3)}}.
 \label{eq:mdot-special-exact-app}
\end{equation}
The common factor $P_\chi$ makes the critical mechanism especially transparent on this line: the largest physical root of $P_\chi(x_c)=0$ simultaneously sends $\zeta$ to zero and produces the simple pole in the flux, which is then canceled by $\rho_{0,\infty}\propto\zeta$.

For $\ell=\chi$, Eq.~\eqref{eq:criticalquartic} becomes
\begin{equation}
 \mathscr C(x_s;\chi,\chi)=-x_sP_\chi(x_s),
\end{equation}
so the physical root is Eq.~\eqref{eq:Pc0}.  Define
\begin{equation}
 y_s(x_s)\equiv c_{{\rm s},s}^2,\qquad y_c\equiv y_s(x_c).
\end{equation}
For a simple root,
\begin{equation}
 y_s-y_c=y'_{s,c}\delta x_s+O(\delta x_s^2),
 \qquad \delta x_s\equiv x_s-x_c,
\end{equation}
while Eq.~\eqref{eq:cinf-generic-main} gives
\begin{equation}
 \zeta
 =-\frac12\Mcal'_{s,c}\delta x_s+O(\delta x_s^2).
\end{equation}
Eliminating $\delta x_s$ yields
\begin{equation}
 \zeta=\Bcrit_\chi(y_s-y_c)
 +O\!\left[(y_s-y_c)^2\right],
 \qquad
 \Bcrit_\chi=-\frac{\Mcal'_{s,c}}{2y'_{s,c}},
\end{equation}
which is Eq.~\eqref{eq:Bspecial}.  Evaluating the derivatives with
\begin{equation}
 y_s(x_s)=\frac{x_s-\chi^2}{\chi^2+x_s(2x_s-3)},
 \qquad
 \Mcal_s(x_s)=\frac{x_s^2}{x_s^2-2x_s+\chi^2}[1-3y_s(x_s)]
\end{equation}
and then imposing $P_\chi(x_c)=0$ gives the coefficient for any chosen $\chi$.  In particular,
\begin{equation}
 \Bcrit_{\chi=0}=\frac{9}{16},
 \qquad
 \Bcrit_{\chi=1}=\frac34.
\end{equation}
These values characterize only the local sonic response on the analytic $\ell=\chi$ slice.  The scalar-mass matching itself follows directly from $\zeta(z,m)$ in Eq.~\eqref{eq:zeta-m-redshift}, without introducing a second near-critical parameter.

\subsection{Exact $\ell=0$ branch}

For $\ell=0$,
\begin{equation}
 \mathscr C=(x_s^2-\chi^2)(x_s^2-6x_s+\chi^2).
\end{equation}
The factor $x_s^2-\chi^2$ lies at or inside the relevant horizon/capture region, while the outer critical root is Eq.~\eqref{eq:ell0root}.  Using $\chi^2=6x_c-x_c^2$ in Eq.~\eqref{eq:css-generic-main} gives
\begin{equation}
 c_{{\rm s},c}^2=\frac1{x_c+3}.
\end{equation}
Substituting into Eq.~\eqref{eq:Psi-generic} yields exactly unity, proving Eq.~\eqref{eq:Psiell0}.  The near-critical derivative coefficient is also analytic:
\begin{equation}
 \Bcrit_{\chi,0}
 =\frac{(x_c-3)^2(x_c+3)}{x_c(x_c^2-3x_c+6)}.
 \label{eq:B-ell0}
\end{equation}

\subsection{Schwarzschild-limit consistency with our previous work}

The Kerr formulas also make two typographical inconsistencies in the supplementary material of our previous work transparent.  Separately, the revised ambient-density normalization discussed in Sec.~\ref{sec:zeta-microphysics} corrects an astrophysical normalization assumption rather than an algebraic error in the Bondi solution.  Their sonic relation itself implies
\begin{equation}
 c_{{\rm s},c}^2=\frac19 \quad\Longrightarrow\quad r_{s,c}=6M,
\end{equation}
not $24M$.  This is exactly the $\chi\to0$ limit of Eq.~\eqref{eq:Pc0}.  Likewise, if one translates back to the near-critical parameter $\epsilon_{\rm old}$ used in our previous Schwarzschild supplement, where $\zeta\simeq(9/16)\epsilon_{\rm old}$, then $r_B=2M/\zeta$ gives
\begin{equation}
 r_B\simeq\frac{32}{9}\frac{M}{\epsilon_{\rm old}},
 \qquad
 r_B\simeq16\frac{\rhob}{\rhodm}M,
 \label{eq:rB-corrected-schwarz}
\end{equation}
rather than the numerical prefactors printed in their Eqs.~(S28) and (S31).  These corrections do not affect the universal SIDM Bondi critical rate $\dot M_{{\rm Bondi},{\rm crit}}=64\pi\rhob M^2$, which follows directly from the cancellation of the near-critical pole against the asymptotic density.

\subsection{Marginally bound boundaries}

For the prograde boundary define $\eta_+\equiv\sqrt{1-\chi}$.  Then
\begin{equation}
 \ell_+=2(1+\eta_+),\qquad x_{\rm mb}^+=(1+\eta_+)^2.
\end{equation}
Direct substitution into Eq.~\eqref{eq:criticalquartic} and Eq.~\eqref{eq:ndef} gives
\begin{equation}
 \mathscr C(x_{\rm mb}^+;\chi,\ell_+)=0,
 \qquad n(x_{\rm mb}^+;\chi,\ell_+)=0.
\end{equation}
The retrograde proof is identical with $\eta_-\equiv\sqrt{1+\chi}$,
\begin{equation}
 \ell_-=-2(1+\eta_-),\qquad x_{\rm mb}^-=(1+\eta_-)^2.
\end{equation}
For nonextremal Kerr, the physical critical branch approaches these roots with $c_{{\rm s},c}^2\to0$ and $\Psic\to0$.  The extremal prograde endpoint is degenerate because the marginally bound orbit coincides with the horizon; the open-domain statement~\eqref{eq:ell-domain} avoids this endpoint subtlety.

\bibliography{Kerr_SIDM_Critical_Bondi_Paper_Draft}

@article{Bondi1952,
 author={Bondi, Hermann}, title={On Spherically Symmetrical Accretion},
 journal={Mon. Not. Roy. Astron. Soc.}, volume={112}, pages={195--204}, year={1952},
 doi={10.1093/mnras/112.2.195}}

@article{Michel1972,
 author={Michel, F. Curtis}, title={Accretion of matter by condensed objects},
 journal={Astrophys. Space Sci.}, volume={15}, pages={153--160}, year={1972}, doi={10.1007/BF00649949}}

@article{Bardeen1972,
 author={Bardeen, James M. and Press, William H. and Teukolsky, Saul A.},
 title={Rotating Black Holes: Locally Nonrotating Frames, Energy Extraction, and Scalar Synchrotron Radiation},
 journal={Astrophys. J.}, volume={178}, pages={347--369}, year={1972}, doi={10.1086/151796}}

@article{Chakrabarti1996,
 author={Chakrabarti, Sandip K.}, title={Accretion processes on a black hole}, journal={Phys. Rept.}, volume={266}, pages={229--390}, year={1996}, doi={10.1016/0370-1573(95)00057-7}}

@article{Goswami2007,
 author={Goswami, Sanghamitra and Khan, Saba Nashreen and Ray, Arnab K. and Das, Tapas Kumar},
 title={Axisymmetric black hole accretion in the Kerr metric as an autonomous dynamical system},
 journal={Mon. Not. Roy. Astron. Soc.}, volume={378}, pages={1407--1417}, year={2007}, doi={10.1111/j.1365-2966.2007.11869.x}}

@article{Das2012,
 author={Das, Tapas K. and Nag, Sankhasubhra and Hegde, Swathi and Bhattacharya, Sourav and Maity, Ishita and Czerny, Bozena and Barai, Paramita and Wiita, Paul J. and Karas, Vladimir and Naskar, Tapan},
 title={Black Hole spin dependence of general relativistic multi-transonic accretion close to the horizon},
 journal={arXiv e-prints}, eprint={1211.6952}, archivePrefix={arXiv}, year={2012}}

@article{ShaikhDas2018,
 author={Shaikh, Md Arif and Das, Tapas Kumar}, title={Linear perturbations of low angular momentum accretion flow in the Kerr metric and the corresponding emergent gravity phenomena},
 journal={Phys. Rev. D}, volume={98}, pages={123022}, year={2018}, doi={10.1103/PhysRevD.98.123022}}

@article{DasNandi2019,
 author={Das, Santabrata and Nandi, Anuj}, title={Low angular momentum relativistic hot accretion flow around Kerr black holes with variable adiabatic index},
 journal={Mon. Not. Roy. Astron. Soc.}, volume={484}, pages={3209--3218}, year={2019}, doi={10.1093/mnras/stz168}}

@article{Aguayo2021,
 author={Aguayo-Ortiz, Alejandro and Sarbach, Olivier and Tejeda, Emilio}, title={Choked accretion onto a Kerr black hole},
 journal={Phys. Rev. D}, volume={103}, pages={023003}, year={2021}, doi={10.1103/PhysRevD.103.023003}}

@article{Kimura2021,
 author={Kimura, Masashi and Harada, Tomohiro and Naruko, Atsushi and Toma, Kenji},
 title={Backreaction of Mass and Angular Momentum Accretion on Black Holes: General Formulation of the Metric Perturbations and Application to the Blandford-Znajek Process},
 journal={Prog. Theor. Exp. Phys.}, volume={2021}, pages={093E03}, year={2021}, doi={10.1093/ptep/ptab101}}

@article{Feng2022,
 author={Feng, Wei-Xiang and Parisi, Alessandro and Chen, Chian-Shu and Lin, Feng-Li}, title={Self-interacting dark scalar spikes around black holes via relativistic Bondi accretion},
 journal={JCAP}, volume={08}, pages={032}, year={2022}, doi={10.1088/1475-7516/2022/08/032}}

@article{ChenLin2026,
 author={Chen, Chian-Shu and Lin, Feng-Li}, title={A Unified Dark-Matter--Driven Relativistic Bondi Route to Black-Hole Growth from Stellar to Supermassive Scales},
 journal={Astrophys. J. Lett.}, volume={1001}, pages={L39}, year={2026}, eprint={2511.09311}, archivePrefix={arXiv}, primaryClass={hep-ph}}

@article{Volonteri2010,
 author={Volonteri, Marta}, title={Formation of supermassive black holes}, journal={Astron. Astrophys. Rev.}, volume={18}, pages={279--315}, year={2010}, doi={10.1007/s00159-010-0029-x}}

@article{Inayoshi2020,
 author={Inayoshi, Kohei and Visbal, Eli and Haiman, Zoltan}, title={The Assembly of the First Massive Black Holes},
 journal={Ann. Rev. Astron. Astrophys.}, volume={58}, pages={27--97}, year={2020}}

@article{Maiolino2024,
 author={Maiolino, Roberto and others}, title={A small and vigorous black hole in the early Universe}, journal={Nature}, volume={627}, pages={59--63}, year={2024}}

@article{Bogdan2024,
 author={Bogd\'an, \'Akos and others}, title={Evidence for heavy-seed origin of early supermassive black holes from a z=10 X-ray quasar}, journal={Nature Astron.}, volume={8}, pages={126--133}, year={2024}}

@article{Juodzbalis2024,
 author={Juod\v{z}balis, Ignas and others}, title={A dormant overmassive black hole in the early Universe}, journal={Nature}, volume={636}, pages={594--599}, year={2024}}

@article{PeitzAppl1997,
 author={Peitz, Jochen and Appl, Stefan},
 title={Viscous accretion discs around rotating black holes},
 journal={Mon. Not. Roy. Astron. Soc.}, volume={286}, pages={681--695}, year={1997},
 doi={10.1093/mnras/286.3.681}}

@article{Takahashi2007,
 author={Takahashi, Rohta},
 title={Horizon-penetrating transonic accretion discs around rotating black holes},
 journal={Mon. Not. Roy. Astron. Soc.}, volume={382}, pages={567--593}, year={2007},
 doi={10.1111/j.1365-2966.2007.11877.x}, eprint={0705.0048}, archivePrefix={arXiv}}

@article{Sukova2017,
 author={Sukova, Petra and Charzynski, Szymon and Janiuk, Agnieszka},
 title={Shocks in the relativistic transonic accretion with low angular momentum},
 journal={Mon. Not. Roy. Astron. Soc.}, volume={472}, pages={4327--4342}, year={2017},
 doi={10.1093/mnras/stx2254}, eprint={1709.01824}, archivePrefix={arXiv}}

@article{Mach2018,
 author={Mach, Patryk and Pirog, Michal and Font, Jose A.},
 title={Relativistic low angular momentum accretion: long time evolution of hydrodynamical inviscid flows},
 journal={Class. Quant. Grav.}, volume={35}, pages={095005}, year={2018},
 doi={10.1088/1361-6382/aab333}, eprint={1803.04032}, archivePrefix={arXiv}}

@article{Dihingia2025,
 author={Dihingia, Indu K. and Uniyal, Akhil and Mizuno, Yosuke},
 title={The Fate of Transonic Shocks around Black Holes and their Future Astrophysical Implications},
 journal={arXiv e-prints}, year={2025}, eprint={2507.23187}, archivePrefix={arXiv}}

@article{Larson2023,
 author={Larson, Rebecca L. and others},
 title={A CEERS Discovery of an Accreting Supermassive Black Hole 570 Myr after the Big Bang: Identifying a Progenitor of Massive z > 6 Quasars},
 journal={Astrophys. J. Lett.}, volume={953}, pages={L29}, year={2023},
 doi={10.3847/2041-8213/ace619}}

@article{Harikane2023,
 author={Harikane, Yuichi and Zhang, Yechi and Nakajima, Kimihiko and Ouchi, Masami and Isobe, Yuki and Ono, Yoshiaki and Hatano, Shun and Xu, Yi and Umeda, Hiroya},
 title={A JWST/NIRSpec First Census of Broad-Line AGNs at z=4--7: Detection of 10 Faint AGNs with $M_{\rm BH}\sim10^6$--$10^8 M_\odot$ and Their Host Galaxy Properties},
 journal={Astrophys. J.}, volume={959}, pages={39}, year={2023},
 doi={10.3847/1538-4357/ad029e}}

@article{Matthee2024,
 author={Matthee, Jorryt and others},
 title={Little Red Dots: An Abundant Population of Faint Active Galactic Nuclei at z $\sim$ 5 Revealed by the EIGER and FRESCO JWST Surveys},
 journal={Astrophys. J.}, volume={963}, pages={129}, year={2024},
 doi={10.3847/1538-4357/ad2345}}

@article{Lambrides2026,
 author={Lambrides, Erini and Larson, Rebecca L. and Garofali, Kristen and others},
 title={The case for super-Eddington accretion in JWST broad-line active galactic nuclei during the first billion years},
 journal={Nature Astron.}, volume={10}, pages={868--879}, year={2026},
 doi={10.1038/s41550-026-02813-w}}

@article{Pariev1996,
 author={Pariev, V. I.},
 title={Hydrodynamic accretion on to a rapidly rotating Kerr black hole},
 journal={Mon. Not. Roy. Astron. Soc.}, volume={283}, pages={1264--1280}, year={1996},
 doi={10.1093/mnras/283.4.1264}}

@article{Font1999,
 author={Font, Jose A. and Ibanez, Jose M. and Papadopoulos, Philippos},
 title={Non-axisymmetric relativistic Bondi--Hoyle accretion on to a Kerr black hole},
 journal={Mon. Not. Roy. Astron. Soc.}, volume={305}, pages={920--936}, year={1999},
 doi={10.1046/j.1365-8711.1999.02459.x}}

@article{Bromm:2002hb,
  author = {Bromm, Volker and Loeb, Abraham},
  title = {Formation of the First Supermassive Black Holes},
  journal = {Astrophys. J.},
  volume = {596},
  pages = {34--46},
  year = {2003},
  eprint = {astro-ph/0212400}
}

@article{Begelman:2006db,
  author = {Begelman, Mitchell C. and Volonteri, Marta and Rees, Martin J.},
  title = {Formation of Supermassive Black Holes by Direct Collapse in Pregalactic Halos},
  journal = {Mon. Not. Roy. Astron. Soc.},
  volume = {370},
  pages = {289--298},
  year = {2006},
  eprint = {astro-ph/0602363}
}

@article{2002RvMP...74.1015W,
  author = {Woosley, S. E. and Heger, A. and Weaver, T. A.},
  title = {The Evolution and Explosion of Massive Stars},
  journal = {Rev. Mod. Phys.},
  volume = {74},
  pages = {1015--1071},
  year = {2002}
}

@article{Madau:2014pta,
  author = {Madau, Piero and Haardt, Francesco and Dotti, Massimo},
  title = {Super-Critical Growth of Massive Black Holes from Stellar-Mass Seeds},
  journal = {Astrophys. J. Lett.},
  volume = {784},
  pages = {L38},
  year = {2014},
  eprint = {1402.6995},
  archivePrefix = {arXiv},
  primaryClass = {astro-ph.CO}
}

@article{Volonteri:2002vz,
  author = {Volonteri, Marta and Haardt, Francesco and Madau, Piero},
  title = {The Assembly and Merging History of Supermassive Black Holes in Hierarchical Models of Galaxy Formation},
  journal = {Astrophys. J.},
  volume = {582},
  pages = {559--573},
  year = {2003},
  eprint = {astro-ph/0207276}
}

@article{Carr:2020xqk,
  author = {Carr, Bernard and Kuhnel, Florian},
  title = {Primordial Black Holes as Dark Matter: Recent Developments},
  journal = {Ann. Rev. Nucl. Part. Sci.},
  volume = {70},
  pages = {355--394},
  year = {2020},
  eprint = {2006.02838},
  archivePrefix = {arXiv},
  primaryClass = {astro-ph.CO}
}

@article{Inayoshi:2019fun,
  author = {Inayoshi, Kohei and Visbal, Eli and Haiman, Zoltan},
  title = {The Assembly of the First Massive Black Holes},
  journal = {Ann. Rev. Astron. Astrophys.},
  volume = {58},
  pages = {27--97},
  year = {2020},
  eprint = {1911.05791},
  archivePrefix = {arXiv},
  primaryClass = {astro-ph.GA}
}

\end{document}